\documentclass{emulateapj} 

\usepackage {afterpage}
\usepackage {float}
\usepackage {placeins}

\gdef\1054{MS\,1054--03}

\def\simgeq{{\raise.0ex\hbox{$\mathchar"013E$}\mkern-14mu\lower1.2ex\hbox{$\mathchar"0218$}}} 

\begin {document}

\title {On Sizes, Kinematics, M/L Gradients, and Light Profiles of Massive Compact Galaxies at $z \sim 2$}

\author{Stijn Wuyts\altaffilmark{1,2}, Thomas J. Cox\altaffilmark{3}, Christopher C. Hayward\altaffilmark{1}, Marijn Franx\altaffilmark{4}, Lars Hernquist\altaffilmark{1}, Philip F. Hopkins\altaffilmark{5}, Patrik Jonsson\altaffilmark{1,2}, Pieter G. van Dokkum\altaffilmark{6}}
\altaffiltext{1}{Harvard-Smithsonian Center for Astrophysics, 60 Garden Street, Cambridge, MA 02138}
\altaffiltext{2}{W. M. Keck Postdoctoral Fellow}
\altaffiltext{3}{Carnegie Observatories, 813 Santa Barbara Street, Pasadena, CA 91101}
\altaffiltext{4}{Leiden University, Leiden Observatory, P.O. Box 9513, NL-2300 RA, Leiden, The Netherlands.}
\altaffiltext{5}{Department of Astronomy, University of California Berkeley, Berkeley, CA 94720}
\altaffiltext{6}{Department of Astronomy, Yale University, New Haven, CT 06520-8101}

\begin{abstract}
We present a detailed analysis of the structure and resolved stellar
populations of simulated merger remnants, and compare them to
observations of compact quiescent galaxies at $z \sim 2$.  We find
that major merging is a viable mechanism to produce systems of $\sim
10^{11}\ M_{\sun}$ and $\sim 1$ kpc size, provided the gas fraction at
the time of final coalescence is high ($\sim 40$\%), and provided that
the progenitors are compact star-forming galaxies, as expected at high
redshift.  Their integrated spectral energy distributions and velocity
dispersions are in good agreement with the observations, and their
position in the $(v_{\rm maj}/\sigma, \epsilon)$ diagram traces the
upper envelope of the distribution of lower redshift early-type
galaxies.  The simulated merger remnants show time- and
sightline-dependent $M/L$ ratio gradients that result from a
superposition of radially dependent stellar age, stellar metallicity,
and extinction.  The median ratio of effective radius in rest-frame
$V$-band light to that in mass surface density is $\sim 2$ during the
quiescent remnant phase.  This is typically expressed by a negative
color gradient (i.e., red core), which we expect to correlate with the
integrated color of the system.  Finally, the simulations differ from
the observations in their surface brightness profile shape.  The
simulated remnants are typically best fit by high ($n \gg 4$) Sersic
indices, whereas observed quiescent galaxies at $z \sim 2$ tend to be less
cuspy ($\langle n \rangle \sim 2.3$).  Limiting early star formation in the
progenitors may be required to prevent the simulated merger remnants
from having extended wings.
\end{abstract}


\section {Introduction}
\label{intro.sec}
Recent surveys of the high-redshift universe have identified a
substantial population of massive quiescent galaxies, already in place
at $z \geq 2$ (e.g., Labb\'{e} et al. 2005; Daddi et al. 2005; van
Dokkum et al. 2006; Kriek et al. 2006).  Studies of their structural
parameters have convincingly ruled out simple monolithic collapse
models, in which little to no structural evolution is expected.
Instead, observational studies find galaxies, and especially quiescent
galaxies, to grow significantly in size as time progresses (e.g.,
Trujillo et al. 2006; Toft et al. 2007; Zirm et al. 2007; van Dokkum
et al. 2008; Cimatti et al. 2008; van der Wel et al. 2008; Franx et
al. 2008; Buitrago et al. 2008).  At $z \sim 2.3$, massive quiescent
galaxies are typically 5 times more compact, and two orders of
magnitude more dense than local ellipticals of the same mass (e.g.,
van Dokkum et al. 2008, hereafter vD08).  Bezanson et al. (2009) note
that, even though their effective densities (measured within one
effective radius $r_e$) are strikingly high compared to local
ellipticals, the central densities measured within a fixed aperture of
1 kpc exceed those of local ellipticals by no more than a factor 2-3
(see also Hopkins et al. 2009d).  This observation suggests an
inside-out growth, in agreement with stacking results by van Dokkum et
al. (2010).

Motivated by these recent observational developments, several
mechanisms have been proposed to incorporate the constraints on
structural evolution into galaxy formation theories.  In most cases,
minor and/or major mergers are invoked to explain the observed size
evolution (Khochfar \& Silk 2006a; Naab et al. 2007, 2009; Hopkins et
al. 2009a).  Briefly, mergers were more gas-rich at high redshifts,
and hence formed a larger fraction of their stars in a nuclear
dissipational component, explaining their compact nature.  Subsequent
(dry) merging activity puffs up the system without adding too much
mass or new stars, which would violate observational constraints.
However, alternative scenarios involving an expansion of the stellar
distribution as response to significant mass losses have been
suggested as well (Fan et al. 2008).  In either case, an accurate
observational characterization of the size-mass relation provides a
crucial test for galaxy formation models.

Given the paucity of kinematic mass measurements based on
absorption-line spectra of $z>1.5$ galaxies (although see Cenarro \&
Trujillo 2009; Cappellari et al. 2009; van Dokkum, Kriek \& Franx
2009), studies of the high-redshift size-mass scaling relation to date
have focussed on stellar mass estimates from spectral energy
distribution (SED) modeling (e.g., Williams et al. 2010).  Significant
systematic uncertainties related to the assumption of an IMF (e.g.,van
Dokkum 2008; Dav\'{e} 2008; Wilkins et al. 2008) and the choice of a
stellar population synthesis code (e.g., Maraston et al. 2006; Wuyts
et al. 2007; Muzzin et al. 2009a) remain.  Uncertainties related to the
star formation history, metallicity, and dust attenuation are
relatively modest for the quiescent population, they only contribute
significantly to the overall uncertainty during earlier, actively
star-forming phases (Wuyts et al. 2009a).  However important an
accurate characterization of mass, we focus in this paper on the
measurement of the second parameter of the scaling relation: galaxy
size.

Observations probe the projected distribution of light, sampling it by
a discrete number of pixels after it was smeared by a point spread
function (PSF).  In addition, the signal is superposed by noise.  The
translation to a physically more meaningful mass profile involves the
assumption of a mass-to-light ratio $M/L$.  Although often for
simplicity assumed to be a constant, spatial variations in $M/L$ may
occur due to age, metallicity and/or dust gradients.  Furthermore,
since the total size of a galaxy is ill-defined, one refers to
(circularized) size as the radius $r_e$ containing half the mass.
Given the finite image resolution, this quantity is generally obtained
by fitting a template profile, taking pixelization and PSF smearing
into account.  In most of the literature, a one-component Sersic
(1968) profile has been adopted, providing satisfyingly flat residual
images given the noise level of the observations.

Numerical simulations provide an excellent tool for the interpretation
of galaxy structure.  The simulated data offers a three-dimensional
view of the mass, age, and metallicity profile at high resolution,
free of sky noise\footnote[1]{The finite number of particles
introduces particle noise, but for the simulations analyzed in this
paper this is negligible on the scales we study.}.  By feeding the
output to a radiative transfer code and producing mock observations,
each of the above aspects related to the nature of observational data
can be isolated, and its effect analyzed.  For example, contrasting
the light profiles of local gas-rich merger remnants and ellipticals
with those of simulated merger remnants, Hopkins et al. (2008b, 2009b)
demonstrated that a two-component profile (consisting of an inner
dissipational, and outer violently relaxed component) provides both a
better fit and a physically more meaningful interpretation of their
structure than a single Sersic profile.

In this paper, we compare the structure of simulated merger remnants
to the best observations of compact quiescent galaxies at $z \sim 2$
to date.  In addition, we discuss the presence of M/L ratio gradients
that may bias measurements of the half-mass radius, and can be
revealed by multi-wavelength structural studies with the
high-resolution cameras onboard {\it Hubble Space Telescope}.  This
study complements the comparison between observations and merger
simulations by Wuyts et al. (2009b) that focussed on the integrated
colors, number and mass densities of high-redshift quiescent galaxies.

We describe the simulations, and mock observations based thereupon in
\S\ref{simulations.sec}.  There, we also caution for artificial
heating in simulating regimes of extreme density.  We discuss the
relation between size and mass for quiescent galaxies with a range of
formation histories in \S\ref{masssize.sec}, and address their
kinematics in \S\ref{sigma.sec}.  Next, we analyze radial variations
in the $M/L$ ratio, their origin in terms of stellar populations, and
color gradients through which they manifest themselves
(\S\ref{MLgradient.sec}).  Finally, we exploit realistic mock
observations of the simulations to fit their surface brightness
profiles alongside real high-redshift compact galaxies.
\S\ref{profiles.sec} highlights the cuspiness of the simulated
profiles, and explores possible origins of an apparent profile
mismatch with respect to the observations.  We summarize the results
in \S\ref{summary.sec}.

Throughout this paper, we adopt the following cosmological parameters:
$(\Omega _M, \Omega _{\Lambda}, h) = (0.3, 0.7, 0.7)$.

\section {The simulations}
\label{simulations.sec}

\subsection {Main characteristics}
\label{sim_main.sec}

\begin {figure}[htbp]
\centering
\plotone{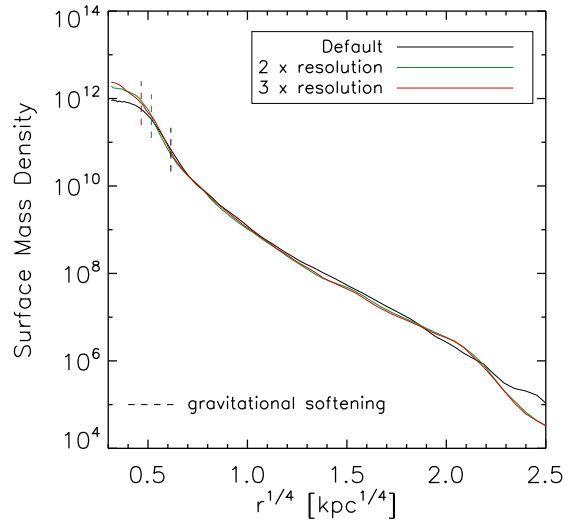}
\caption{Surface mass density profiles of a compact merger remnant
with final stellar mass of $1.7 \times 10^{11}\ M_{\sun}$, simulated
with 1, 2, and 3 times the default spatial resolution adopted in this
paper.  The profiles are converged outside of a softening length.
\label{resol.fig}}
\end {figure}

\begin {figure*}[t]
\centering
\plotone{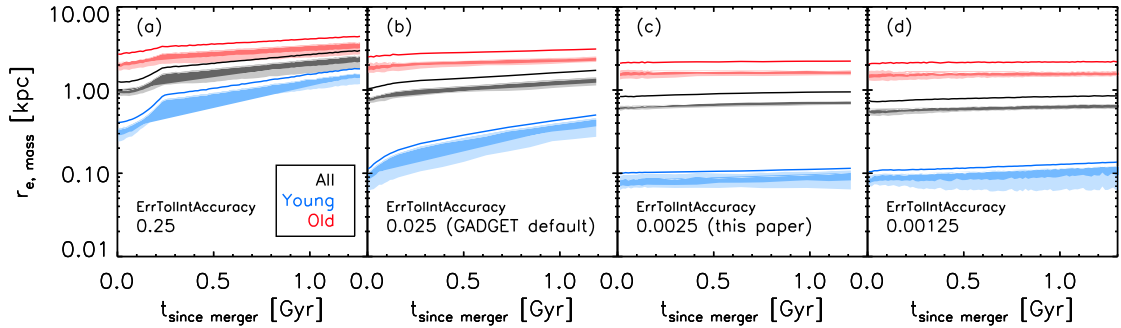}
\caption{Size evolution after final coalescence of a compact merger
remnant with a final stellar mass of $1.7 \times 10^{11}\ M_{\sun}$.
The top line for each color-coded component indicates the 3D half-mass
radius, whereas the light (dark) polygon below illustrates the full
(central 50-percentile) range of 2D projected half-mass radii of the
system as viewed from different sightlines.  Panels (a) to (d)
represent realizations of SPH simulations with increasingly finer
timestep resolution and otherwise identical conditions.  Numerical
heating causes an artificial increase in size when the orbital paths
of SPH particles in a dense environment are integrated with
insufficient resolution.  For a system of $\sim 10^{11}\ M_{\sun}$ and
half-mass radius of $\sim 1$ kpc, the size evolution is not yet
converged for the default GADGET timestep resolution.
\label{timestep.fig}}
\end {figure*}

This work is based on a suite of smoothed particle hydrodynamic (SPH)
simulations representing isolated and merging galaxies at redshifts $z
\sim 0$ to 3 that were performed with the GADGET-2 code (Springel
2005).  The code uses an entropy-conserving formalism of SPH (Springel
\& Hernquist 2002), and includes gas cooling, a multi-phase model for
the interstellar medium (ISM) to describe star formation and supernova
feedback (Springel \& Hernquist 2003), and a prescription for
supermassive black hole growth and feedback (Springel, Di Matteo \&
Hernquist 2005; Di Matteo, Springel \& Hernquist 2005).

The progenitor disks in our simulations are embedded in an extended
dark matter halo with a Hernquist (1990) profile.  They span a range of
baryonic masses from $7 \times 10^{9}\ M_{\sun}$ to $4 \times 10^{11}\
M_{\sun}$, and initial gas fractions of 20 - 80\%.  We mostly focus on
the subset of massive gas-rich merger simulations in which the
progenitors were scaled to approximate the structure of disk galaxies
at redshift $z=3$, following Robertson et al. (2006).  Briefly, this
means that the mass- and redshift-dependent halo concentration
measured by Bullock et al. (2001) was adopted:
\begin {equation}
C_{vir}(M_{vir}, z) \approx 9 \left( \frac{M_{vir}}{M_{coll,0}} \right)^{-0.13} (1+z)^{-1},
\label {Cvir.eq}
\end {equation}
where $M_{coll,0} \sim 8 \times 10^{12}\ h^{-1}\ M_{\sun}$ is the
linear collapse mass at z=0.  The virial mass and virial radius of the progenitors were scaled as follows:
\begin {eqnarray}
M_{vir} = \frac{V_{vir}^3}{10GH(z)} \\
R_{vir} = \frac{V_{vir}}{10H(z)},
\label {Mvir.eq}
\end {eqnarray}
where $V_{vir}$ is the virial velocity and $H(z)$ is the Hubble
parameter.  Disk sizes were initialized according to the Mo et
al. (1998) formalism for dissipational disk galaxy formation assuming
the fraction of the total angular momentum contained in the disk
equals the fraction of the total mass contained in the disk.  The disk
scale length is then derived from the halo concentration $C_{vir}$
(Eq.\ \ref{Cvir.eq}) and the galaxy spin $\lambda$, where we adopt a
default $\lambda=0.033$, as motivated by cosmological N-body
simulations (Vitvitska et al. 2002).  In practice, this means that the
$z=3$-scaled progenitor disks have effective radii that are a factor
1.7 smaller than similar mass galaxies today, in agreement with the
observed size evolution of star-forming galaxies (Franx et al. 2008).

Equal-mass mergers were simulated for a range of orbital
configurations, from coplanar to polar to various tilted disk
orientations, following Cox et al. (2006).  Specifically, we analyzed
simulations with the range of spin axis orientations of the progenitor
disks listed, in standard spherical coordinates, in Table\
\ref{spinorientations.tab}.  In typical runs, each of the two
progenitor galaxies initially consists of 60000 to 120000 dark matter
particles, 40000 gas and 40000 stellar disk particles, and one black
hole sink particle.  For the mass range probed by our simulations,
this corresponds to typical mass resolutions of the baryonic and dark
matter particles of $M_{\rm bar} = 1-8 \times 10^5\ M_{\sun}$ and
$M_{\rm DM} = 0.5-4 \times 10^7\ M_{\sun}$ respectively.In addition to
equal-mass mergers, we also ran simulations in which the disk galaxy
was left to evolve in isolation, simulations of unequal-mass mergers,
and scenarios where a merger remnant undergoes subsequent merging.
\begin{deluxetable}{lccccl}[b]
\tablecolumns{6}
\tablecaption{Disk Orientations
\label{spinorientations.tab}
}
\tablehead{ID\tablenotemark{a} & $\theta _1$ & $\phi _1$ & $\theta _2$ & $\phi _2$ & Comments}
\startdata
h & 0    & 0   & 0   & 0   & both prograde \\
b & 180  & 0   & 0   & 0   & prograde-retrograde \\
d & 90   & 0   & 0   & 0   & polar \\
e & 30   & 60  & -30 & 45  & tilted 1 \\
g & 150  & 0   & -30 & 45  & tilted 2 \\
i & 0    & 0   & 71  & 30  & Barnes orientations\tablenotemark{b} \\
j & -109 & 90  & 71  & 90  & $\Downarrow$ \\
k & -109 & -30 & 71  & -30 & \\
l & -109 & 30  & 180 & 0   & \\
m & 0    & 0   & 71  & 90  & \\
n & -109 & -30 & 71  & 30  & \\
o & -109 & 30  & 71  & -30 & \\
p & -109 & 90  & 180 & 0   & \\
\enddata
\tablenotetext{a}{\scriptsize Unique orientation identification from Cox et al. (2006).}
\tablenotetext{b}{\scriptsize Selected by Barnes (1992) to be unbiased
initial disk orientations according to the coordinates of two
oppositely directed tetrahedra.}
\end{deluxetable}

We adopted a gravitational softening length of $\epsilon = 140$ pc.  A
resolution study for one of our fiducial merger simulations, where we
increased the mass resolution by a factor 4 and repeated the
simulations with two and three times the spatial resolution (i.e.,
adopting softening lengths that are half or a third of our default)
and otherwise identical conditions confirms that the adopted
resolution is sufficient for the purposes of this paper (Figure\
\ref{resol.fig}).  The total stellar mass of the merger remnant varies
by a few percent only when increasing the spatial resolution.  At
radii larger than the softening length, the surface mass density
profiles are converged at the 10\% level.  Within a radius of 140 pc,
the amount of stellar mass assembled in the nucleus tends to be
somewhat larger in the higher resolution runs.  We furthermore refined
the GADGET-2 timestep resolution by a factor $\sqrt{10}$ with respect
to the GADGET-2 default, resulting in runtimes that are longer by the
same factor.  Figure\ \ref{timestep.fig} illustrates the importance of
the latter choice when studying (sub)structure at high spatial
densities.  The panels illustrate four realizations of a gas-rich
merger simulation with a final stellar mass of $1.7 \times 10^{11}\
M_{\sun}$ and half-light radius of $\sim 1$ kpc, identical except for
the timestep resolution with which particle orbits are integrated.
The timestep scales with $\sim \sqrt{\epsilon \eta}$, where the
respective value of the GADGET-2 parameter $\eta$ = ErrTolIntAccuracy
is shown in the panel.  For each realization, we show the 3D (top
line) and 2D (bottom polygon) half-mass radius evolution as function
of time since the merger, which we define as the radius of a sphere
(respectively circle) encompassing half of the (projected) stellar
mass.  Throughout this paper, we define time of merging as the moment
when the peak in star formation activity is reached.  Different colors
represent the size evolution of the galaxy as a whole (black), the
young component of stars that formed within a 250 Myr interval around
final coalescence (blue), and the old component of stars that were
already formed prior to the nuclear starburst (red).  Several
conclusions can be drawn from Figure\ \ref{timestep.fig}.  First,
numerical heating can lead to artificial growth of the simulated
merger remnant.  Since this effect is not uniform over the galaxy, but
manifests itself particularly in the dense and young central
component, it can also artificially alter the profile shape.
Furthermore, the stellar mass formed during the nuclear starburst
steadily decreases when a finer timestep resolution is adopted, by
26\% when comparing simulation (a) to (d) from Figure\
\ref{timestep.fig}, and by 10\% when comparing simulation (b) to (d).
This trend suggests that, in addition to artificial growth during the
post-merger phase, the implementation of star formation and/or
feedback processes may also depend on the length of the integration
time step.  This result has important implications for studies of
structural evolution of massive galaxies based on cosmological
simulations, where finite computational power imposes a delicate
tradeoff between the box size (essential for reliable number
statistics at the massive end), the spatial resolution, and the
timestep resolution.  Clearly, by compromising the timestep resolution
to a level where no numerical convergence is reached, spurious growth
in size and smoothing of cusps will be superposed on galaxy growth by
real physical processes.  

A second conclusion to draw from Figure\ \ref{timestep.fig} is that,
once a converging timestep resolution is adopted, we find no evidence
for size evolution after final coalescence in our binary merger
simulations\footnote[2]{Note that this simulation does not include
subsequent merging or gas infall, and only treats stellar mass loss in
an instantaneous manner (see Springel \& Hernquist 2003).}.  Third,
the young dissipational component in this simulation that started with
a gas fraction of 80\% and was still very gas-rich ($f_{\rm gas} =
0.43$) at the time of final coalescence is more than an order of
magnitude smaller than the old component of previously formed stars.
Finally, by definition the 2D half-mass radius (which depends on
viewing angle) is smaller than the 3D half-mass radius.  We find a
typical $r_{e, 3D} / r_{e, 2D}$ ratio of 1.4.  Furthermore, we find
that the ratio $r_{e, 3D} / r_{e, 2D}$ depends more strongly on
viewing angle for the young component than for the old component,
implying that the recent, dissipational star formation event took
place in a disk-like structure, whereas the violently relaxed
component has a more spherical shape.  We find similar properties for
the merger remnants produced by simulations that started with
different orbital configurations.

\subsection {Translating simulations to observables}
\label{sim_phot.sec}

The fluxes, colors and light profiles of the merger remnants are
computed from the simulation output in two steps.  First, the
intrinsic stellar emission is derived from a stellar population
synthesis code, where we treat each particle as a Simple Stellar
Population (SSP) with its stellar mass, age and metallicity computed
by the GADGET-2 code.  We adopt a Kroupa (2001) IMF and compute the
photometry using Bruzual \& Charlot (2003) or Maraston (2005) models.
We find that our results are independent of the choice of stellar
population synthesis code.  Initial conditions (stellar age and
metallicity) of the stellar and gas particles present at the start of
the simulation were set by a simple closed box formalism detailed by
Wuyts et al. (2009a).  The precise choice of these initial stellar
population properties has a negligible impact on the nature of the
merger remnants that are the focus of this paper.

In step two, we use the information on the gas distribution and
enrichment to compute the attenuating effect of dust on the emerging
galaxy light.  We perform our analysis using two independent codes
that are frequently used in the literature.  The first is a
line-of-sight attenatuation code (LOS) that simply computes the
wavelength-dependent absorption to each stellar particle from the
metallicity-weighted diffuse gas column density between that stellar
particle and the observer (see Hopkins et al. 2005; Robertson et
al. 2007; Wuyts et al. 2009a,b).  The second code, SUNRISE (Jonsson
2006; Jonsson, Groves \& Cox 2010), uses a Monte Carlo methodology to
track photon packets on their way through the dusty ISM, and models
the effects of both absorption and scattering.  In addition, SUNRISE
uses a sub-grid model to account for the attenuation by birthclouds
(HII and photodissociation regions) that surround young star clusters,
computed using the photoionization code MAPPINGS (Groves et al. 2008).
Furthermore, SUNRISE tracks dust temperatures and re-emission of
absorbed light at longer wavelengths (see, e.g., Younger et al. 2009;
Narayanan et al. 2010).  However, since no high-resolution structural
information is available in the far-infrared, and dust re-emission is
negligible during the quiescent merger remnant phase, this aspect of
the radiative transfer is not discussed in this paper.  Likewise, the
GADGET-2 simulation keeps track of gas accretion onto a central
supermassive black hole, but its emission is negligible during the
quiescent merger remnant phase of interest.

We compute realizations of the radiative transfer with attenuation
laws representative of the Milky Way and the Small Magellanic Cloud.
Unless specifically mentioned, our conclusions are qualitatively the
same independent of the radiative transfer code or attenuation law
used.

\subsection {Mock images}
\label{sim_mock.sec}

We use the vD08 sample of compact quiescent galaxies at $z \sim 2.3$
as reference.  All 9 galaxies have a spectroscopically confirmed
Balmer/4000\AA\ break (Kriek et al. 2006).  Seven of them were
observed for 3 orbits in the F160W filter using the NIC2 camera on
{\it HST}, probing their rest-frame optical emission.  The two
brighest galaxies were exposed for 2 orbits.  At a drizzled pixel scale of
0.0378'', the point-spread function (PSF) is well sampled, and the
observations yielded the highest-resolution surface brightness maps of
such galaxies to date.

In order to establish a fair comparison, we produce mock observations
placing the simulated merger remnants at the same redshift, using the
same observed filter and an identical pixel scale.  We convolved the
resulting postage stamps with a PSF extracted from the vD08 NIC2
images, applied Poisson noise, and added them into empty regions of
the NIC2 images in order to guarantee similar noise properties.  We
randomized the sub-pixel position of the simulated galaxies, but find
that the recovered structural properties are independent of this
treatment.

In the following, we first analyze the {\it true} half-mass
(\S\ref{masssize.sec}) and half-light (\S\ref{MLgradient.sec}) radii
of the simulated galaxies as measured from the full (noise-free and
high-resolution) information available from simulation and radiative
transfer.  Next, we include the effects of pixelization, finite
resolution, and limited signal-to-noise by running the two-dimensional
fitting code GALFIT (Peng et al. 2002) on the real and mock NIC2
images (\S\ref{profiles.sec}).

\section {Simulated merger remnants on the size - mass relation}
\label{masssize.sec}

\begin {figure*}[t]
\plotone{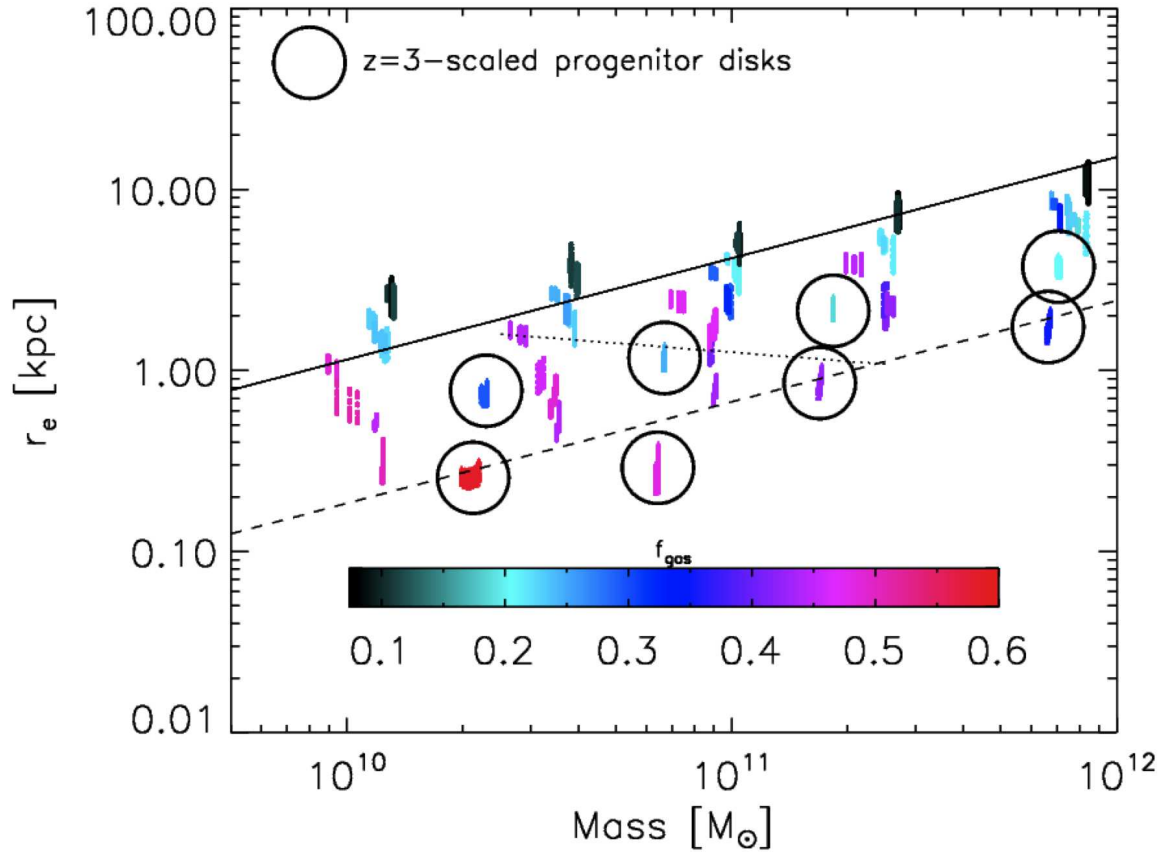}
\caption{Size-mass diagram of simulated merger remnants, color-coded
by the gas fraction of the merger from which they descend, measured
125 Myr prior to the peak in star formation rate.  Simulations where
the progenitors were scaled to represent disk galaxies at $z=3$ are
encircled.  The solid line indicates the size-mass relation of
present-day early-type galaxies (Shen et al. 2003).  The quiescent
galaxies at $z \sim 2.3$ from vD08 lie on the dashed line.  The dotted
line indicates the size-mass relation at $z \sim 2.3$ based on an
extrapolation of the mass-dependent size evolution measured by Ryan et
al. 2010.  At a given mass, merger remnants are more compact as the
dissipational fraction increases.
\label{sizemass.fig}}
\end{figure*}
Wuyts et al. (2009b) demonstrated that the integrated stellar
population properties (rest-frame optical and optical-to-NIR colors,
as well as specific star formation rates) of simulated merger remnants
are consistent with those of observed quiescent galaxies at $z \sim
2$.  Here, we investigate whether a major merger formation scenario
can also account for the remarkable compactness of the observed
quiescent systems.  A useful diagnostic to address this question is
the size-mass relation.  In Figure\ \ref{sizemass.fig}, we mark the
local spheroid size-mass relation (Shen et al. 2003) with a solid
line.  As reliable size measurements of complete samples of
high-redshift quiescent galaxies are only available over a narrow mass
range, we indicate a fiducial size-mass relation at redshift $z \sim
2.3$ (dashed line) by assuming the $z \sim 0$ slope and adopting the
median size and mass of the vD08 sample as zero-point anchor.  We note
that both observational (Trujillo et al. 2006; Ryan et al. 2010) and
theoretical (Khochfar \& Silk 2006a) studies have suggested that the
size-mass relation becomes shallower with redshift.  Ryan et
al. (2010) report a mass dependence of the power-law index $\alpha$ in
the size evolution $R_e/R_{e,z=0} = (1+z)^{- \alpha}$ of $\alpha
\approx -1.8 + 1.4 \log (M_* / 10^9\ M_{\sun})$.  We caution that
this fit is driven by measurements of galaxies at $0<z<2$, and, as
stated by Ryan et al. (2010), should be considered preliminary at
best.  Nevertheless, it is illustrative to plot the corresponding
size-mass relation at $z \sim 2.3$ (dotted line).  We conclude that
the ultra-compactness of massive ($> 10^{11}\ M_{\sun}$) quiescent
galaxies at $z \sim 2$ is a robust result.  Tighter constraints on the
sizes of high-redshift spheroids of lower mass will be essential to
understand whether they formed through similar mechanisms.  The main
focus of this paper is on compact galaxies at the high-mass end.

Overplotted in Figure\ \ref{sizemass.fig}, we show the 2D
(i.e., projected) half-mass radii of simulated merger remnants as a
function of their stellar mass, as seen from 100 viewing angles
uniformly spread over a sphere.  We consider merger simulations with a
range of stellar masses, and in each case only plot the snapshots more
than 100 Myr after the final starburst.  Encircled are the remnants
produced by merging progenitors that were scaled to represent $z = 3$
disks (i.e., reaching the remnant phase around $z \sim 2.3$).  The
other simulations started out with progenitors scaled to represent
local disk galaxies.  For every simulation, we define a dissipational
fraction $f_{\rm gas}$ as the gas fraction of the system 125 Myr before
the peak in star formation rate (SFR) is reached.  Even though some of
our simulations start with initial gas fractions as high as 80\%,
$f_{\rm gas}$ rarely reaches values above 50\%, because rapid star
formation in the progenitor disks, especially during first passage,
consumes significant amounts of gas before final coalescence.

Figure\ \ref{sizemass.fig} clearly illustrates that, at a given mass,
the size of a merger remnant is smaller when the dissipational
fraction is higher, and even more so when the progenitor disks had a
more compact nature to start with.  Tidal torques are responsible for
channeling large amounts of gas to the central region (Barnes \&
Hernquist 1991, 1996) where it is consumed in a starburst (Mihos \&
Hernquist 1994, 1996).  We conclude that the location of the observed
$z \sim 2$ quiescent galaxies on the size-mass diagram (the vD08
sample has median properties $r_e = 0.9$ kpc, $M = 1.7 \times 10^{11}\
M_{\sun}$) can straightforwardly be explained by major merger activity
provided the progenitors at high redshift were more gas-rich and had
scalelengths smaller than today's disk galaxies.  This idea was first
formulated by Khochfar \& Silk (2006a), and is further detailed by
Hopkins et al. (2009a; 2010a).  Significant obervational support for
increased gas fractions in star-forming galaxies towards higher
redshift was inferred from H$\alpha$ spectroscopy by Erb et
al. (2006), and more recently confirmed on the basis of molecular line
measurements (Baker et al. 2004; Coppin et al. 2007; Daddi et
al. 2008; Tacconi et al. 2010).  Likewise, a decrease in the size of
star-forming galaxies at a given mass is observationally well
established (Trujillo et al. 2006; Franx et al. 2008; Williams et
al. 2010).

\section {Kinematics}
\label{sigma.sec}

\begin {figure}[htbp]
\plotone{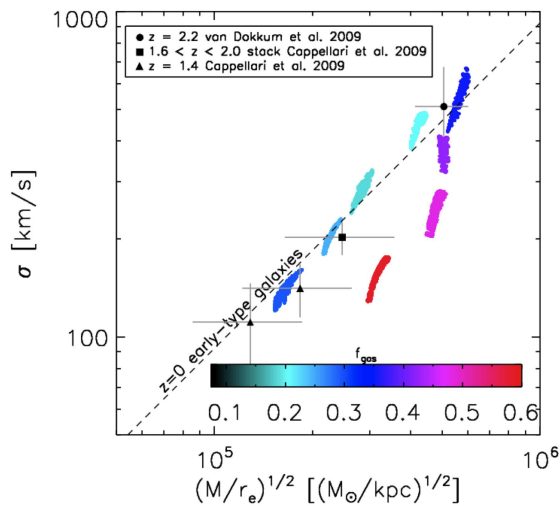}
\caption{Velocity dispersion of quiescent galaxies at $1.4<z<2.2$ as 
function of the virial estimator $\sqrt{\frac{M}{r_e}}$ based on their
stellar mass and size.  Simulated merger remnants span a similar
range, but at a given $\sigma$ extend to higher values of the virial
estimator when the dissipational fraction is large ($> 40$\%).
\label{sigma.fig}}
\vspace{-0.2in}
\end{figure}
A proper characterization of galaxy structure requires knowledge of
the mass-to-light ratio to determine the stellar mass, and of its
gradient to translate the measured half-light radius to a radius
containing half the mass (see \S\ref{MLgradient.sec}).  One way to
circumvent $M/L$ ratio effects, is probing the central potential
directly through kinematics.  However, measuring the velocity
dispersions of quiescent $z \sim 2$ galaxies from their absorption
line spectra is extremely expensive in terms of telescope time.  To
date, one such measurement (van Dokkum, Kriek \& Franx 2009) at $z >
2$ has been carried out, totaling 29 hours on a 8m class telescope.
At somewhat lower redshift and mass ($\log M = 10 - 11$), a stacked
measurement of similar red nuggets at $1.6<z<2.0$, as well as two
individual measurements at $z \sim 1.4$ were presented by Cappellari
et al. (2009).  In Figure\ \ref{sigma.fig}, we compare the measured
velocity dispersion to the virial estimator $\sqrt{\frac{M}{r_e}}$,
where $M$ is the stellar mass derived from SED modeling.  The dashed
line indicates the proportionality followed by present-day early-type
galaxies (van Dokkum \& Stanford 2003; Cappellari et al. 2006).
Color-coded by their gas fraction shortly before the final starburst,
we overplot the same simulated merger remnants whose progenitors were
scaled to represent high-redshift star-forming galaxies as displayed
in Figure\ \ref{sizemass.fig}.  Both the observed and simulated
quiescent galaxies show a clear correlation between the measured
velocity dispersion and what would be estimated based on virial
arguments.  Within the error bars, the observational results are all
consistent with having the same scaling between $\sqrt{\frac{M}{r_e}}$
and $\sigma$ as early-type galaxies in the nearby universe.
Simulations of gas-rich mergers are able to produce remnants with
similar velocity dispersions.  At a given velocity dispersion, the
virial estimator $\sqrt{\frac{M}{r_e}}$ is larger for runs with a
higher dissipational fraction, implying that mergers of varying
dissipational fraction are non-homologous (see also Robertson et
al. 2006; Hopkins, Cox \& Hernquist 2008).  Briefly, the scaling
factor $k$ in the relation
\begin{equation}
\sigma = k \sqrt{\frac{M}{r_e}},
\end{equation}
where $M$ is the stellar mass and $r_e$ the stellar half-mass radius,
depends on the profile shape of the stellar mass distribution (lower
$k$ for cuspier systems), and on the baryon-to-dark matter ratio
within the stellar effective radius (lower $k$ for more
baryon-dominated centers).  Our simulations show that both factors are
a strong function of the dissipational fraction $f_{\rm gas}$ of the
merger.  Systems with larger $f_{\rm gas}$ are increasingly
baryon-dominated in their centers.  Their cuspy profile shapes are
discussed at length in Section\ \ref{cuspiness.sec}.

\begin {figure}[t]
\plotone{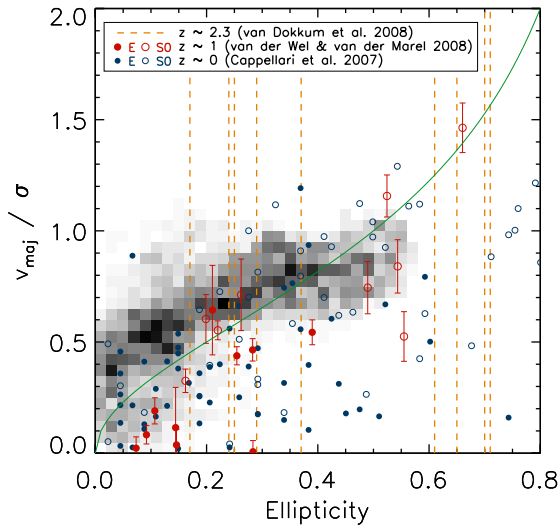}
\caption{(v/$\sigma$, $\epsilon$) diagram for simulated merger
remnants of $\sim 10^{11}\ M_{\sun}$ and $\sim 1$ kpc size ({\it
gray-coded distribution}).  Ellipticities of quiescent galaxies at $z
\sim 2.3$ with similar mass and size from vD08 are indicated with
dashed vertical lines.  Early-type galaxies at $z \sim 0$ and $z \sim
1$ are overplotted for reference.  The green line marks the relation
expected for an oblate isotropic rotator.  The simulations suggest
that rotation may on average play a more dominant role in early-type
galaxies at high redshift than at low redshift, due to the
increasingly dissipational processes through which they formed.
\label{vsig_eps.fig}}
\end{figure}
Although such measurements are not yet available observationally at $z
\sim 2$, it is interesting to consider the degree of rotation expected
in high-redshift compact quiescent galaxies.  To this end, we study
the ratio of rotational over random motion ($\frac{v_{\rm
maj}}{\sigma}$, where the rotation velocity is measured along the
major axis) as a function of ellipticity in Figure\
\ref{vsig_eps.fig}.  The gray-coded distribution indicates the locus
occupied by quiescent merger remnants that match the vD08 galaxies in
size and mass.  These remnants are the product of gas-rich merger
simulations with a range of orbital configurations, and we observe
them from 100 sightlines uniformly distributed over a sphere.  The
green line marks the relation expected for an oblate isotropic rotator
(Binney 1978):

\begin{equation}
\frac{v}{\sigma} = \sqrt{\frac{\epsilon}{1 - \epsilon}}
\end{equation}

Vertical dashed lines indicate the ellipticities of the $z \sim 2.3$
compact galaxies of vD08, as measured on the high-resolution NIC2
images.  For reference, we also plot the location of massive
early-type galaxies (of larger size than those at $z \sim 2$) at $z
\sim 1$ (van der Wel \& van der Marel 2008) and at $z \sim 0$ (a
complete sample extracted from the HyperLeda database by Paturel et
al. 2003, as detailed by van der Wel \& van der Marel 2008).  The
closed and open circles represent galaxies with an E and S0 morphology
respectively.

The rotation parameter of the simulated merger remnants increases with
ellipticity.  At a given ellipticity, the compact remnants show a
larger degree of rotational support than the bulk of early-type
galaxies today.  Their kinematics resemble those of the fastest
rotators of the $z \sim 0$ sample at each ellipticity.  Cox et
al. (2006) demonstrated that the rotation parameters and flattening of
local spheroids with less rotation can be reproduced by merger
simulations tuned to lower redshift, with lower gas fractions (see
also Naab, Jesseit \& Burkert 2006).  A hint of increasing rotational
support towards higher redshifts, where mergers are expected to be
increasingly dissipational, may already be observed when comparing the
measurements at $z \sim 1$ to those at $z \sim 0$.  It remains to be
seen whether this trend extends to $z \sim 2$, and whether similarly
high values of $\frac{v}{\sigma}$, of up to unity, are present as
anticipated by our simulations.  The large range of ellipticities in
the vD08 sample, reaching values of $\epsilon > 0.6$ which are larger
than found in our simulations, may point to rotation being indeed
important in these systems.

\section {$M/L$ ratio gradients}
\label{MLgradient.sec}

Size measurements have traditionally been carried out using one
waveband only, typically the longest wavelength available at high
resolution.  By definition, this technique probes a monochromatic
surface brightness profile.  Evidently, the half-light radius derived
from it equals the physically more meaningful half-mass radius only
under the assumption of a spatially constant $M/L$ ratio.

\begin {figure}[t]
\plotone{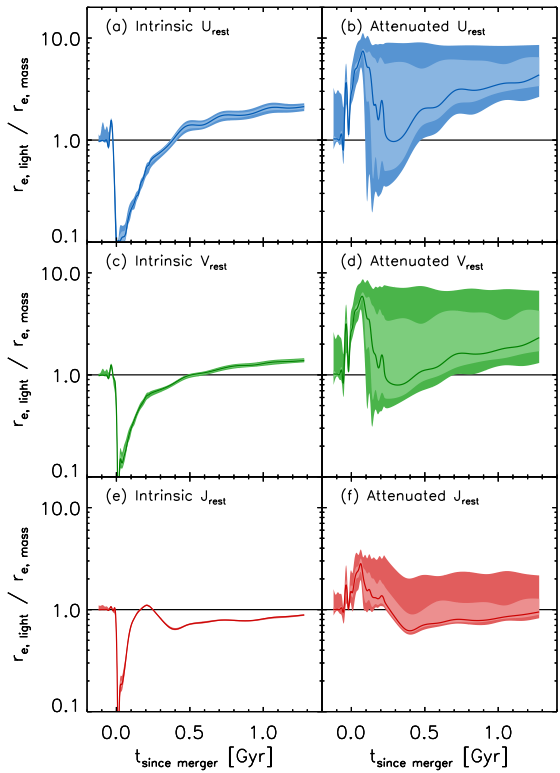}
\caption{\small Ratio of half-light to half-mass radius as function of time
since the merger, as determined from the intrinsic (unattenuated)
photometry ({\it left panels}) and the attenuated photometry ({\it
right panels}).  Mass-to-light ratio gradients originate due to a
combination of age, metallicity, and extinction gradients that depend
on time, wavelength, and viewing angle.  The half-light radius of
quiescent merger remnants measured in the rest-frame $V$-band is
typically larger by a factor 1.5 to 2 than the projected radius
containing half the mass.  This ratio increases to a factor 2-3 in the
rest-frame $U$-band.  Our result implies that the observed compact
quiescent galaxies (see, e.g., vD08), if formed by a similar gas-rich
merger process, may be even more compact in terms of mass than
previously assumed.
\label{Rlight_Rmass_LOS.fig}}
\end {figure}

\begin {figure}[t]
\plotone{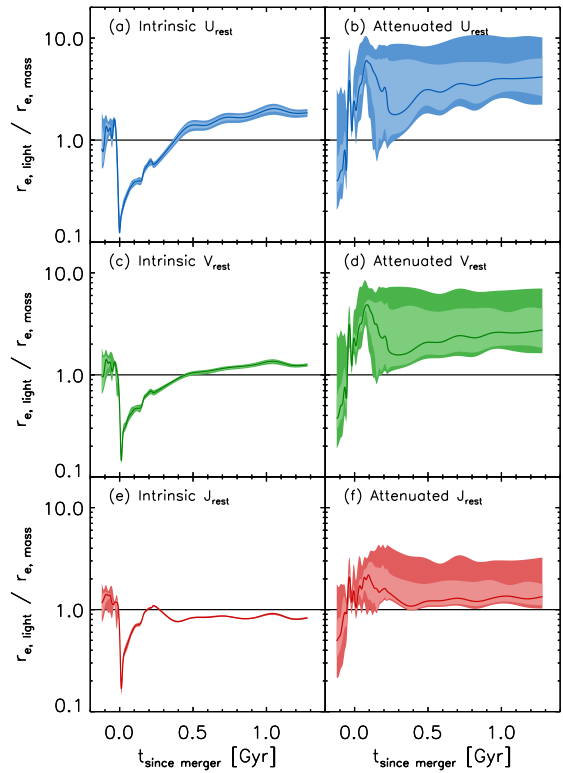}
\caption{
Idem as Figure\ \ref{Rlight_Rmass_LOS.fig}, but with photometry
computed with the SUNRISE radiative transfer code.
\label{Rlight_Rmass_SUNRISE.fig}}
\end {figure}

Having established that gas-rich mergers are a viable mechanism to
collect $\sim 10^{11}\ M_{\sun}$ of baryonic material in a $\sim 1$
kpc radius, we now use our simulations to test this assumption for the
quiescent remnants.  Using the radiative transfer methods described in
\S\ref{sim_phot.sec}, we compute the half-light radius in three
rest-frame wavebands $U$, $V$, and $J$, and contrast it with the
half-mass radius of the system for 100 lines of sight.  Figures\
\ref{Rlight_Rmass_LOS.fig} and\ \ref{Rlight_Rmass_SUNRISE.fig} show
the resulting light-to-mass size ratio $r_{e, \rm{light}} / r_{e,
\rm{mass}}$ as a function of time since the merger for the same
simulation shown in Figure\ \ref{timestep.fig}(c).  The central line
indicates the median evolution over all viewing angles, whereas the
light and dark shaded regions mark the central 50\% and 100\% percentiles
respectively.

\begin {figure}[t]
\plotone{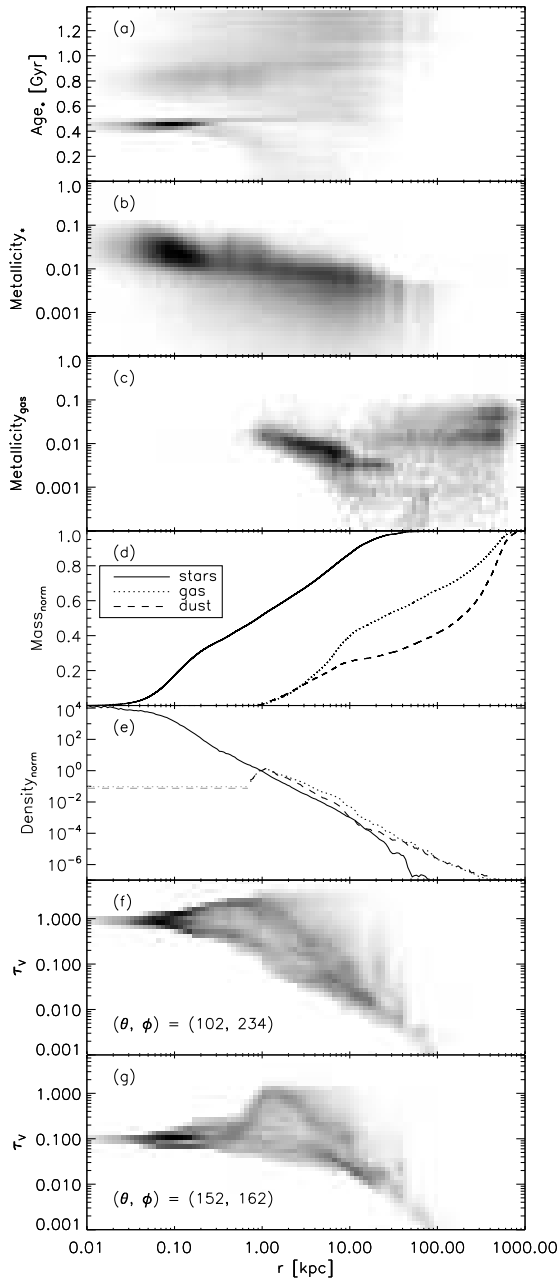}
\caption{\small Radial stellar population profiles of a merger simulation
snapshot during the spheroid phase, half a gigayear after final
coalescence.  We show (a) the distribution of stellar age as a
function of distance to the galaxy center, (b) the stellar metallicity
increasing towards the center, (c) the radial dependence of the
gas-phase metallicity, (d) the normalized stellar, gas and dust mass
distributions, (e) their respective density profiles normalized to the
density at $r = 1$ kpc, (f) and (g) the distribution of optical depths
towards all stellar particles for two random viewing angles.
\label{physgradient.fig}}
\end {figure}

Panels on the left illustrate the time evolution of $r_{e, \rm{light}}
/ r_{e, \rm{mass}}$ ignoring attenuation by dust.  In this case, all
scatter at a given time is due to projection effects.  The scatter is
limited to a few percent in rest-frame $V$, which is the band most
closely corresponding to F160W at $z \sim 2.3$.  Initially, as many
new stars are formed in the nuclear starburst, the intrinsic light
profile is more centrally concentrated than the stellar mass profile
(i.e., $r_{e, \rm{light}} / r_{e, \rm{mass}} \ll 1$).  On timescales
of a few hundred Myr, as the bright and short-lived O and B stars die
out, the nuclear emission dims with respect to the outer parts of the
galaxy, and the $M/L$ gradient flattens.  In fact, around 400 Myr
after the peak in star formation rate, the light profile in the
rest-frame optical becomes more extended than the mass profile.  This
trend is more pronounced as we consider shorter wavelengths, reaching
$r_{e, \rm{light}} / r_{e, \rm{mass}} \sim 2$ in rest-frame $U$.  The
reason for this reversed $M/L$ gradient is that the young stars in the
center formed out of more metal-enriched gas than the stars that
constitute most of the galaxy outskirts, making them fainter and
cooler (Binney \& Merrifield 1998, and references therein).  We
illustrate the physical conditions underlying the $M/L$ ratio
gradients in Figure\ \ref{physgradient.fig} for the snapshot 480 Myr
after the merger.  At this time, the remnant has a specific star
formation rate ($3.7 \times 10^{-11}\ {\rm yr}^{-1}$) and broad-band
colors similar to those of observed quiescent galaxies.  Plotting
stellar age as function of distance from the stellar particles to the
center (Figure\ \ref{physgradient.fig}a), three loci in age can be
distinguished, corresponding to stars formed in the progenitor disks,
during first passage, and during final coalescence.  The radial
position in the merger remnant is clearly correlated with the epoch at
which the stars were formed.  Figure\ \ref{physgradient.fig}a also
illustrates that the low level trickle of star formation present after
the nuclear starburst happens mostly outside the stellar half-mass
radius ($r_{e, 3D} = 0.9$ kpc).  This is a secondary effect
contributing to the larger extent of the light profile compared to the
mass profile.  Figure\ \ref{physgradient.fig}b demonstrates the
presence of a negative stellar metallicity gradient, of slope
$\frac{\Delta \log(Z)}{\Delta \log(r/r_e)} = -0.35$.  This slope is a
factor 1.5 to 2.5 steeper than that of typical massive early-type
galaxies today (Rawle, Smith \& Lucey 2010; Kuntschner et al. 2010),
although those exhibit a large scatter in gradients.  Together with
their compact nature (\S\ref{masssize.sec}) and fast rotation
(\S\ref{sigma.sec}), this implies the gas-rich merger remnants at $z
\sim 2$ cannot evolve passively into present-day massive ellipticals.
Subsequent (dry) merging has been proposed as a mechanism to slow the
rotation, grow the size, and dilute the metallicity gradient, with the
extent of the dilution dependent on the properties of the merger (Di
Matteo et al. 2009).

In addition to stellar age and metallicity, the gas and dust
distribution and extinction column towards stellar particles also vary
radially.  The remaining gas in the $1 < r < 30$ kpc range has a
negative metallicity gradient (Figure\ \ref{physgradient.fig}c), while
at even larger radii the SPH gas particles show a wide spread in
metallicities.  The normalized cumulative distribution of dust (taken
to be proportional to the metallicity-weighted gas distribution) is
contrasted with that of stars and gas in Figure\
\ref{physgradient.fig}d.  The corresponding spherically averaged
density profiles, normalized to the density at 1 kpc, are presented in
Figure\ \ref{physgradient.fig}e.  The gas is distributed over scales
that are an order of magnitude larger than that of the stellar
distribution.  Both gas and dust ($\sim$ metallicity-weighted gas)
reach a peak density near the stellar half-mass radius ($r_{e, 3D} =
0.9$ kpc).  The tail of young stars at $\sim 1$ kpc visible in Figure\
\ref{physgradient.fig}a shows that it is also here that the small
amount of star formation that is still present after the merger
($\frac{SFR}{M} = 3.7 \times 10^{-11}\ $yr$^{-1}$) takes place.  Gas
that during earlier phases resided within this radius has largely been
consumed by star formation, or has been blown out by feedback
processes from supernovae and AGN.  Although this is not visible in
the spherically averaged density profile of Figure\
\ref{physgradient.fig}d, the central cavity devoid of gas and dust is
not spherically symmetric, but rather axisymmetric.  The resulting
radial distribution of optical depths towards the stellar particles
therefore depends on the line of sight.  We illustrate this
distribution for two characteristic viewing angles in panel (f) and
(g).  For the viewing angle presented in panel (f), the effective
attenuation of the region within the half-mass radius is larger than
that outside.  Consequently, the attenuated $V$-band half-light radius
viewed from this angle is 4.8 times larger than the half-mass radius.
For the viewing angle of panel (g), the opposite is the case and the
resulting attenuated $V$-band half-light radius is smaller than the
half-mass radius by a factor 1.4.

The right-hand panels of Figure\ \ref{Rlight_Rmass_LOS.fig} illustrate
how the superposition of a dust, age, and metallicity gradient
translates into the light-to-mass size ratio $r_{e, \rm{light}} /
r_{e, \rm{mass}}$, and how this quantity depends on time relative to
the merger, line of sight, and wavelength.  During the nuclear
starburst, the $M/L$ ratio gradient is completely reversed with
respect to the intrinsic (unattenuated) $M/L$ ratio gradient, because
of central dust obscuration.  The median light-to-mass size ratio then
drops, reaching $r_{e, \rm{light}} / r_{e, \rm{mass}}$ values below 1
around 300 Myr after the peak in star formation rate, as the
central dust content decreases while the young, massive stars in the
center are still alive.  Finally, as the system reaches a quiescent
remnant phase ($t_{\rm since} \gtrsim 500$ Myr), the age gradient has
faded and the combination of a negative extinction and metallicity
gradient results in typical half-light radii of $2.0_{-0.6}^{+2.9}$,
$1.1_{-0.2}^{+2.1}$, and $0.7_{-0.0}^{+0.3}$ times the half-mass
radius at $t_{\rm since} = 500$ Myr in rest-frame $U$, $V$, and $J$
respectively.  At $t_{\rm since} = 1$ Gyr, the half-light radii in the
rest-frame $U$-, $V$-, and $J$-band are $3.4_{-0.6}^{+2.0}$,
$1.8_{-0.3}^{+1.9}$, and $0.8_{-0.0}^{+0.3}$ times the half-mass
radius.  Here, the error bar indicate the central 50\% percentile of
the scatter due to line-of-sight variations.  Typically, the
distribution shows an extended tail towards large $r_{e, \rm{light}} /
r_{e, \rm{mass}}$ ratios.  Both the line-of-sight scatter and the
median value of $r_{e, \rm{light}} / r_{e, \rm{mass}}$ increase
towards shorter wavelengths.

Computing the synthetic photometry with the independent radiative
transfer code SUNRISE (Figure\ \ref{Rlight_Rmass_SUNRISE.fig}) gives
results that are qualitatively consistent with those obtained from the
LOS code.  SUNRISE predicts typical light-to-mass size ratios of 3.6,
2.4, and 1.3 in rest-frame $U$, $V$, and $J$ respectively during the
quiescent phase.  The sightline dependence shows a similar behavior as
seen in the results from the LOS code, i.e., with a tail towards large
$r_{e, \rm{light}} / r_{e, \rm{mass}}$ ratios (ratios of 5 and above).

We tested the dependence of our results on the adopted attenuation
law, on the assumed age and metallicity of stars present at the start
of the simulation, and whether an age and/or metallicity gradient was
already in place in the initial progenitor disks, and find consistent
results with variations much smaller than those from sightline to
sightline.  Likewise, similar results were obtained from simulations
where the progenitor disks were merged with a different orbital
configuration.

Peirani et al. (2010) simulate minor mergers between an elliptical
galaxy and a satellite (spiral galaxy), and find qualitatively similar
stellar age gradients: the young stars that formed by dissipational
processes reside in the center of the remnant.  The evolution of
half-light radii presented by Peirani et al. (2010) resembles that
seen in intrinsic light in our major merger simulations.  However, the
dust columns, and therefore the impact of attenuation on the observerd
$M/L$ gradients, are larger for the major than for the minor merger
simulations.  The assumption of a fixed dust-to-gas rather than
dust-to-metal ratio, and the absence of AGN feedback in the Peirani et
al. (2010) simulations may also contribute to differences in the
predicted $M/L$ gradients.

\begin {figure}[t]
\plotone{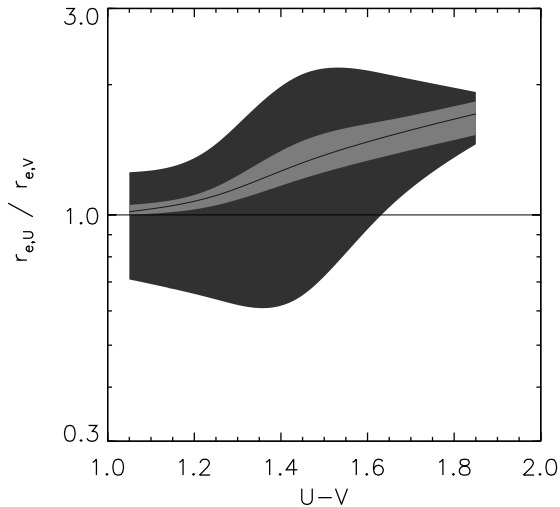}
\caption{ Ratio of $U_{\rm rest}$ to $V_{\rm rest}$ half-light radius
as function of integrated color.  The central line indicates the
median relation, whereas the light- and dark-gray polygon indicate the
50\% and 100\% percentiles of distribution over 100 lines-of-sight
uniformly spread over a sphere.
\label{grad_integrat.fig}}
\end {figure}
In summary, we find that the interplay between an age, metallicity,
and dust gradient manifests itself as a $M/L$ ratio gradient that
depends on time, line of sight and wavelength.  Since an observer can
only probe the latter parameter directly, the detection of color
gradients showing preferentially red cores in compact quiescent
galaxies would support the presented scenario.  Moreover, from the
simulations we expect the internal color gradients to be correlated
with the integrated rest-frame optical color (see Figure\
\ref{grad_integrat.fig}).  We stress that the presence of a red core
does not necessarily imply inside-out growth.  In fact, in this
scenario, and in the absence of subsequent merging (see
\S\ref{subsequent.sec}), the bulk of the youngest stars resides in the
center, but the effect of the age gradient on the internal color
profile is compensated by the presence of metallicity and extinction
gradients.  Finally, if negative $M/L$ ratio gradients are indeed
present, our results may imply that the observed quiescent galaxies
may be an order of magnitude ($\sim 2^3$) more dense than previously
inferred from HST/NICMOS observations (vD08).

\section {Surface brightness profiles}
\label{profiles.sec}

\subsection {The cuspiness of simulated merger remnants}
\label{cuspiness.sec}

\begin {figure*}[htbp]
\epsscale{0.95}
\centering
\plotone{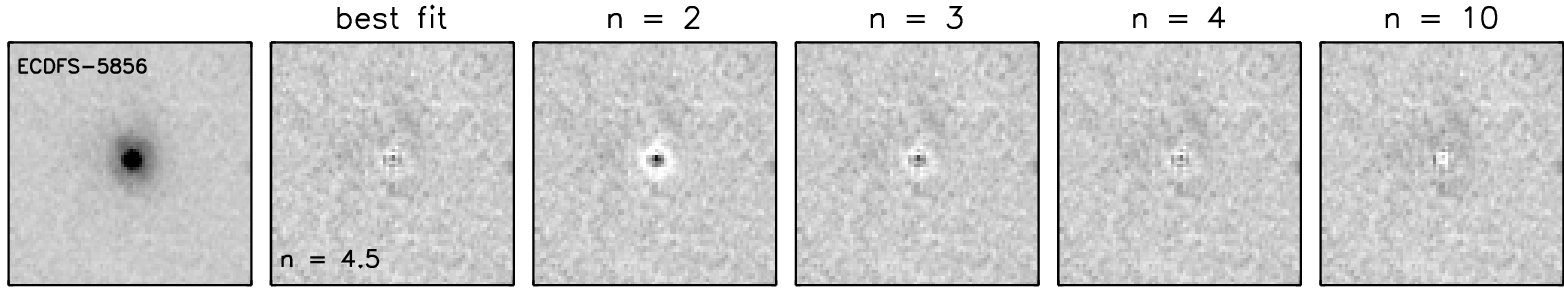}
\vspace{-0.1in}
\plotone{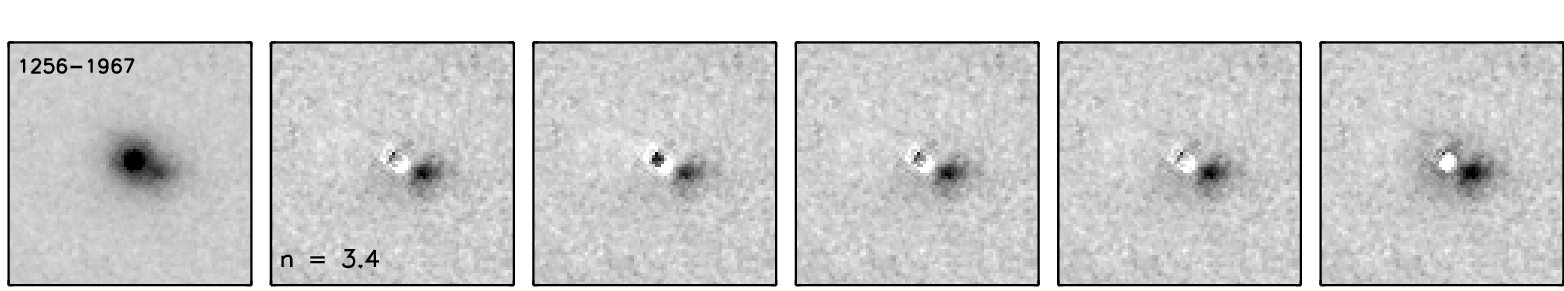}
\vspace{-0.1in}
\plotone{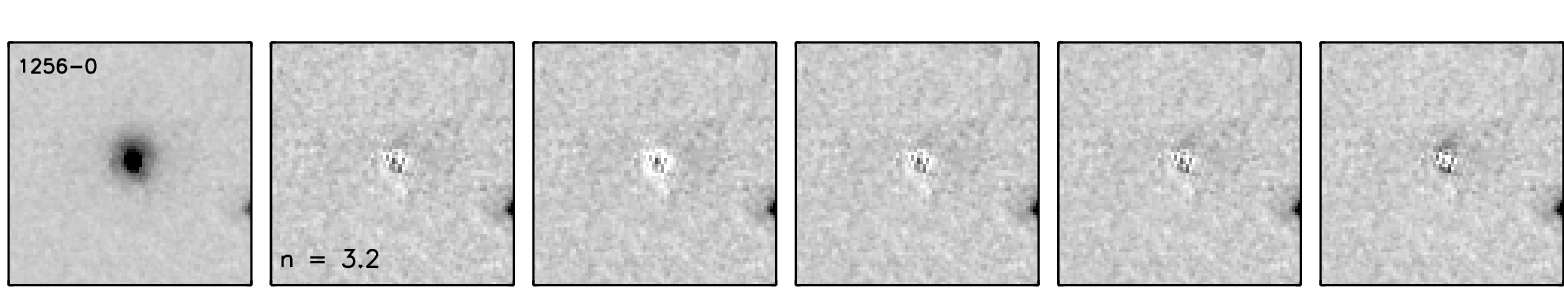}
\vspace{-0.1in}
\plotone{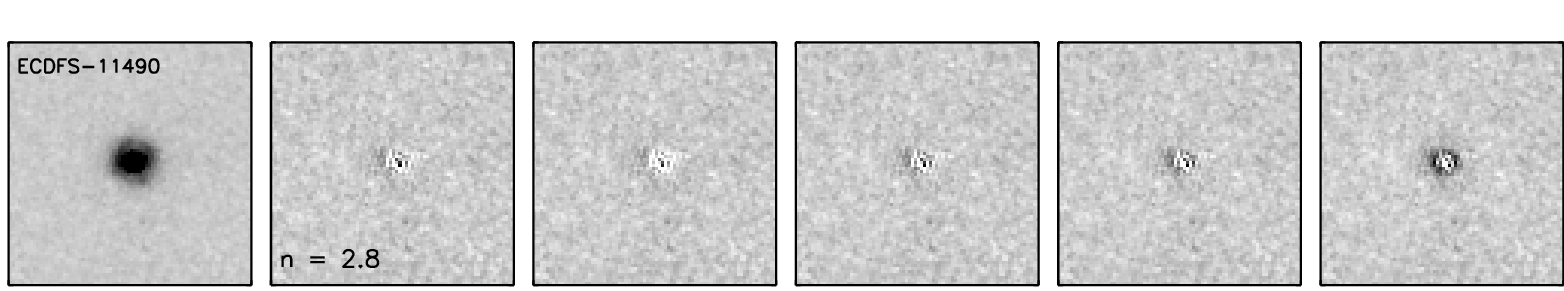}
\vspace{-0.1in}
\plotone{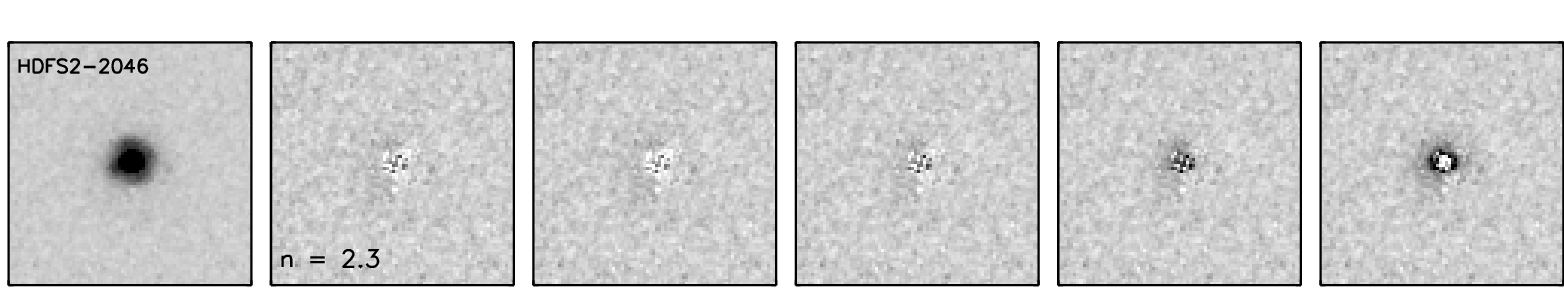}
\vspace{-0.1in}
\plotone{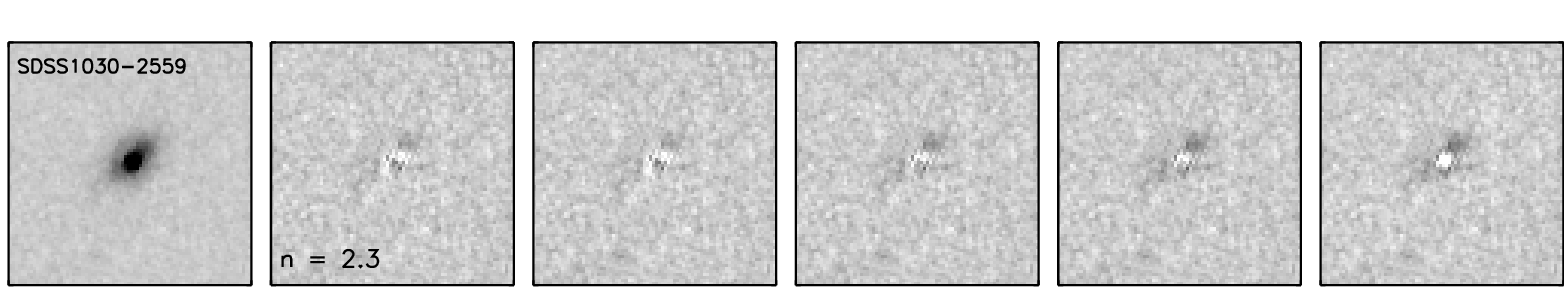}
\vspace{-0.1in}
\plotone{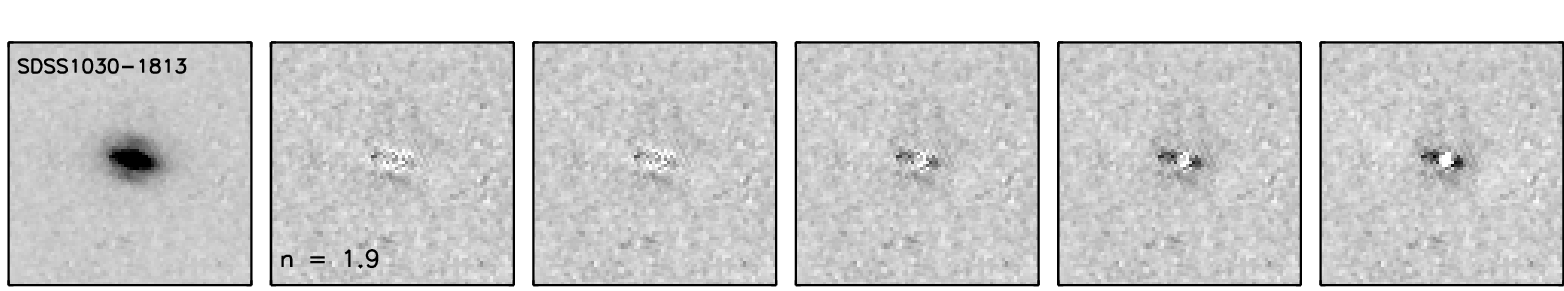}
\vspace{-0.1in}
\plotone{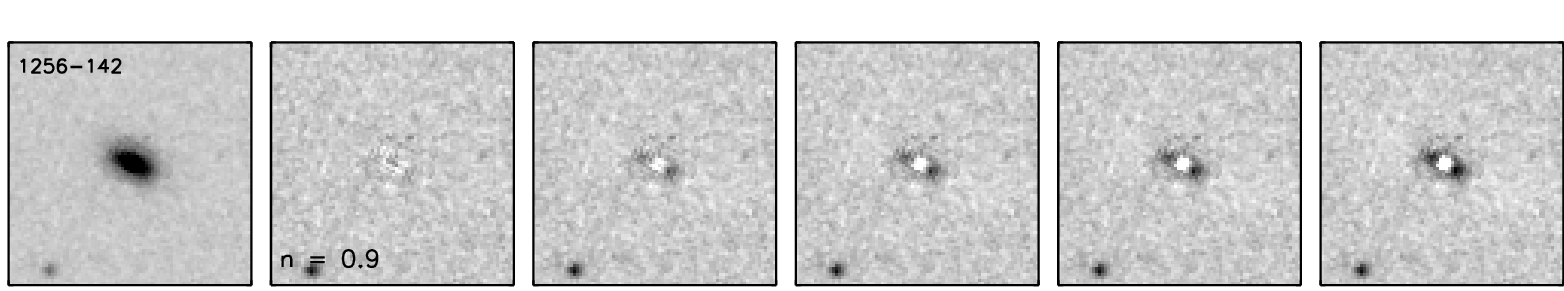}
\vspace{-0.1in}
\plotone{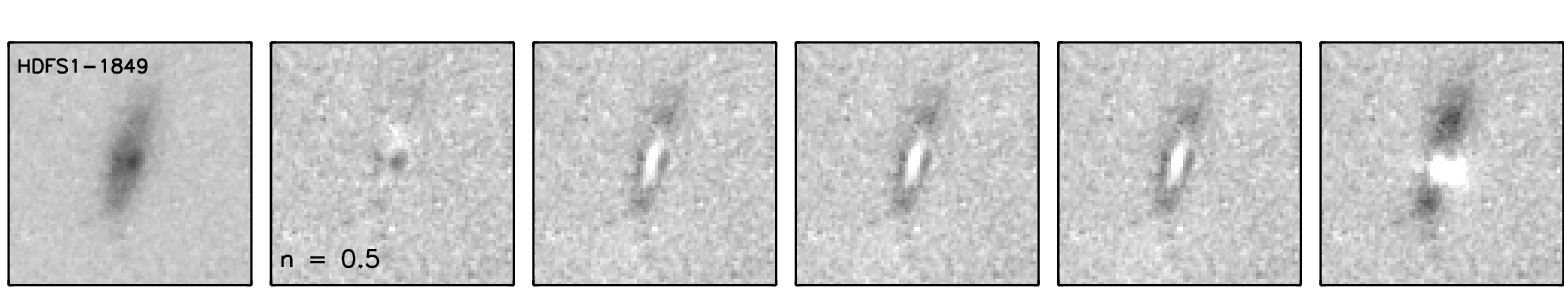}
\epsscale{1}
\caption{NIC2 F160W postage stamp images of the vD08 sample of quiescent
galaxies, and residual images after subtracting the best fit when
leaving $n$ free, or fixing it to $n-2$, 3, 4, and 10 respectively.
The panels are sized 3'' $\times$ 3'', and for clarity the contrast of
the residual images is double that of the observed image.  Typically,
the observed quiescent galaxies are best represented by Sersic
profiles with $n < 4$.
\label{vD08_stamps.fig}}
\end {figure*}

\begin {figure*}[htbp]
\epsscale{0.95}
\centering
\plotone{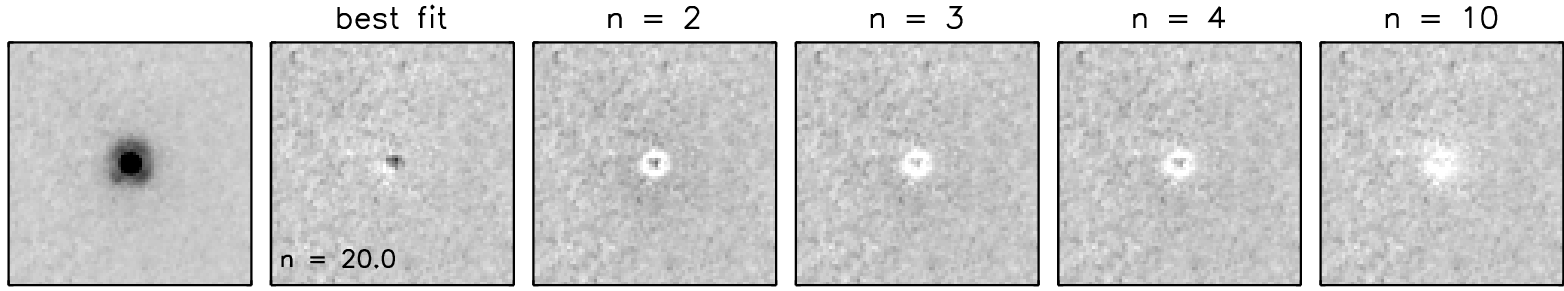}
\vspace{-0.1in}
\plotone{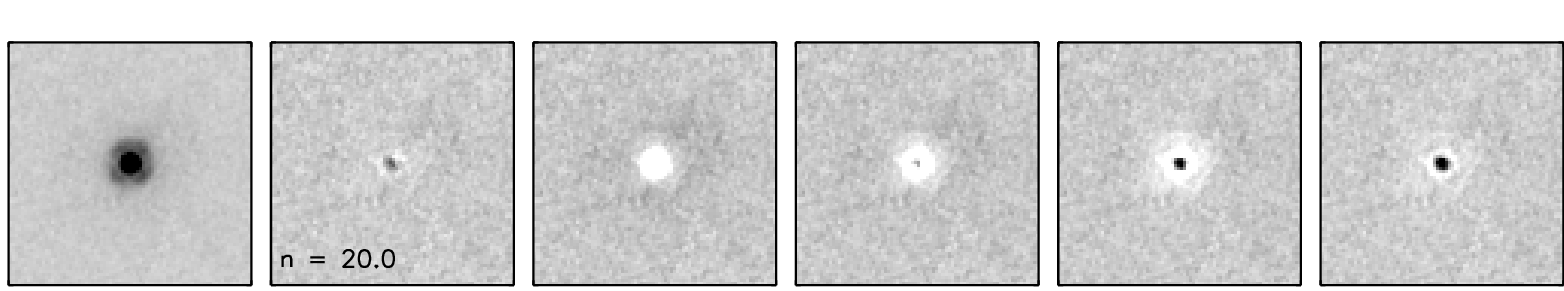}
\vspace{-0.1in}
\plotone{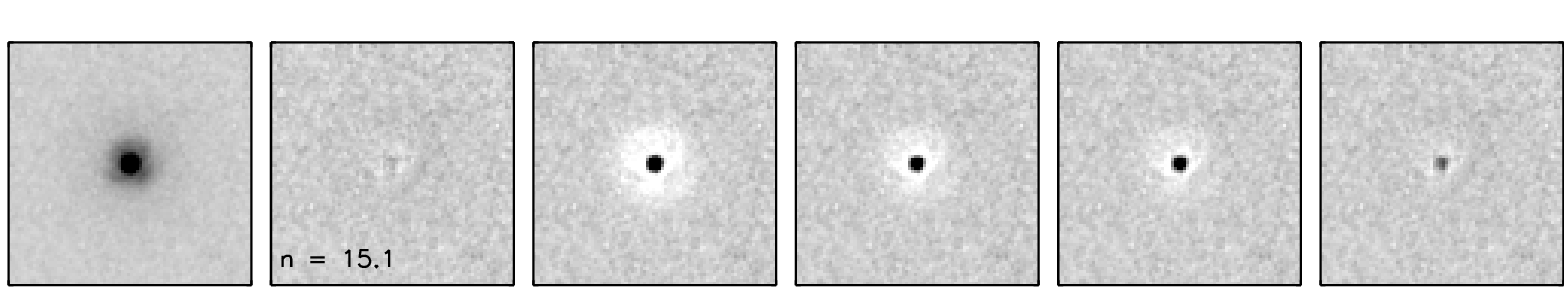}
\vspace{-0.1in}
\plotone{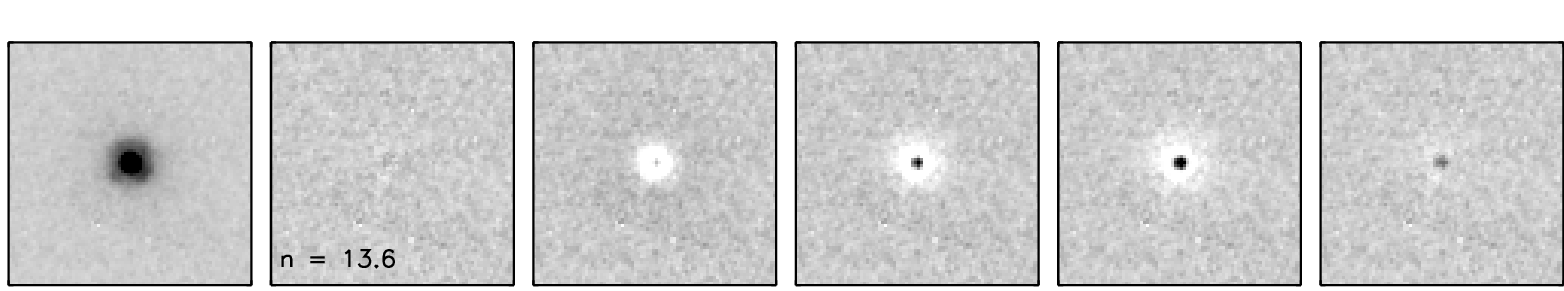}
\vspace{-0.1in}
\plotone{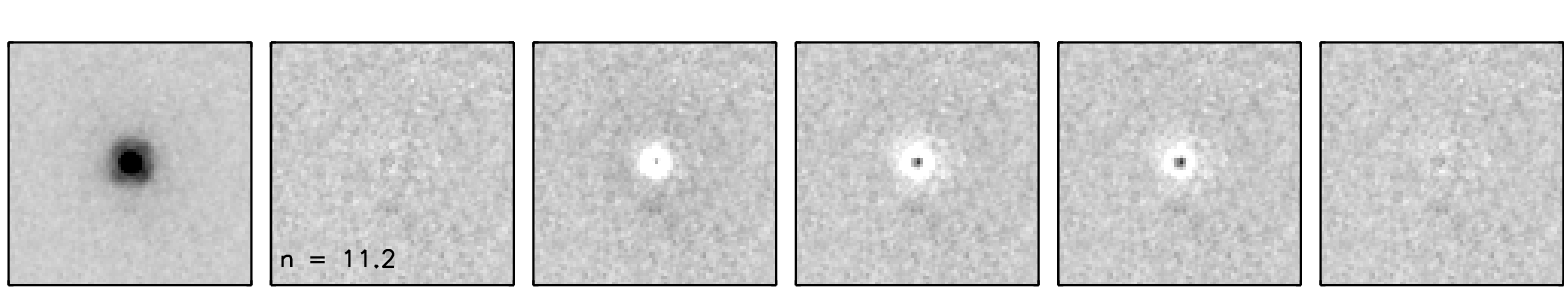}
\vspace{-0.1in}
\plotone{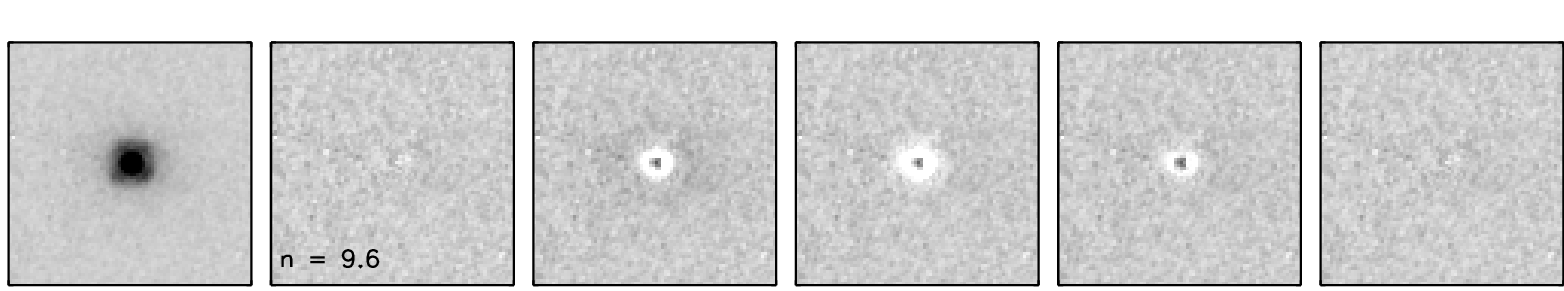}
\vspace{-0.1in}
\plotone{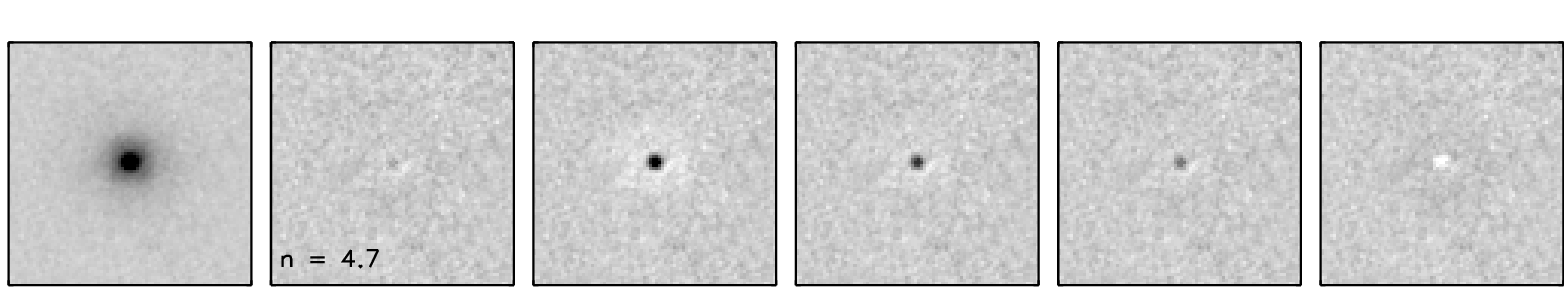}
\vspace{-0.1in}
\plotone{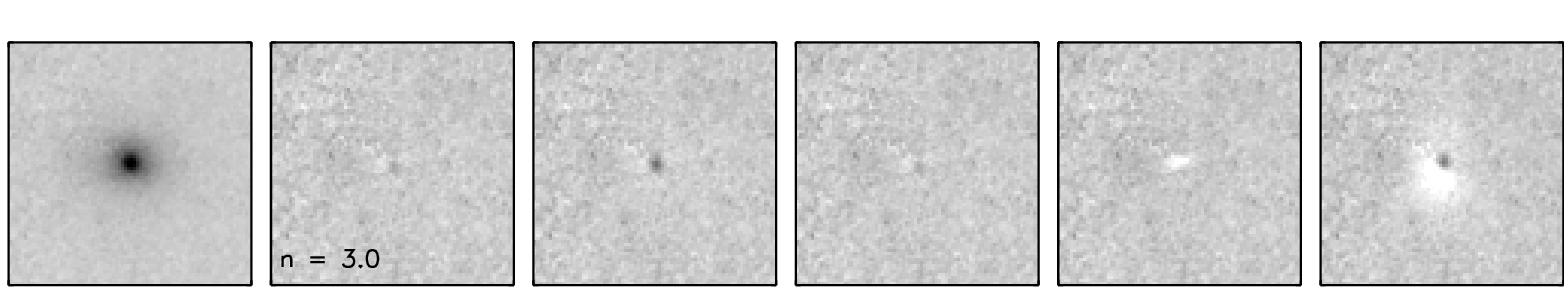}
\vspace{-0.1in}
\plotone{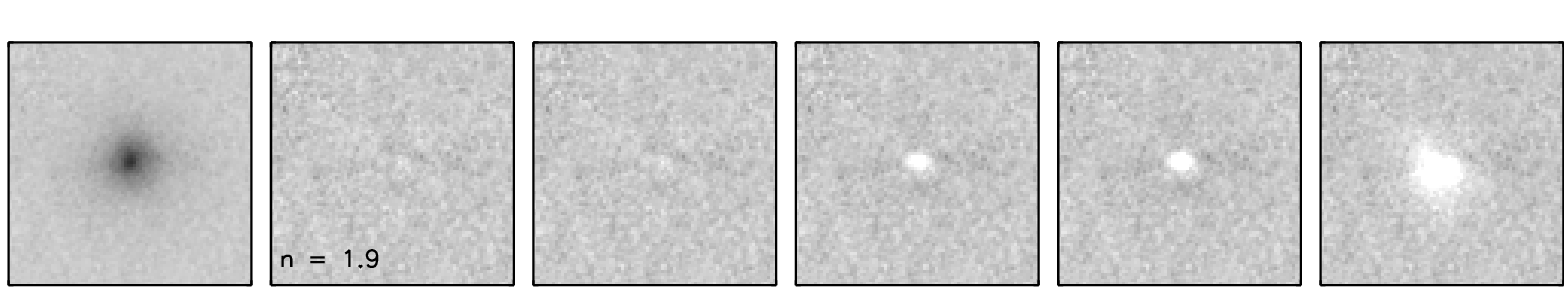}
\epsscale{1}
\caption{\small Mock NIC2 F160W observations of simulated merger remnants.
The panels are 3'' on a side, corresponding to 24.6 kpc at $z=2.3$.
For 80\% of the sightlines, the surface brightness profile is best fit
by a cuspy profile (fit runs into the upper bound of $n=10$).
Significant residuals remain when fixing the Sersic index to lower
values.  For about one fifth of the sightlines (illustrated in the
bottom 3 rows) lower $n$ are obtained, but the recovered half-light
radius for these sightlines is larger than that of observed quiescent
galaxies by a factor of $\sim 5$.
\label{sim_stamps.fig}}
\end {figure*}
We established that gas-rich mergers are a viable mechanism to
assemble dense stellar systems of $\sim 10^{11}\ M_{\sun}$ and $\sim
1$ kpc size (\S\ref{masssize.sec}), and that such a scenario would
leave a color and $M/L$ ratio gradient as imprint
(\S\ref{MLgradient.sec}).  Now, we go beyond the zeroth-order
structural measurement of size and compare how the surface brightness
profiles of simulated merger remnants compare to those of observed
quiescent galaxies at $z \sim 2$.

For this purpose, we use the mock F160W observations described in
\S\ref{sim_mock.sec} that are matched in terms of PSF, pixel size,
wavelength, redshift, and noise properties to the NIC2 observations
of high-redshift spheroids by vD08.  We analyze the real and mock
observations in concert with the two-dimensional fitting code GALFIT
(Peng et al. 2002) using identical settings.  In Figure\
\ref{vD08_stamps.fig}, we show postage stamps of the real NIC2
images by vD08.  Next to each image, we plot the residual images after
subtracting the best-fit Sersic profile when leaving the Sersic index
$n$ free, or fixing it to $n=2$, $n=3$, $n=4$, and $n=10$
respectively.  We rank the objects by best-fit Sersic index and find
they span a large range from $n=4.5$ to $n=0.5$, reproducing the vD08
results.  In most cases, fixing the Sersic index to $n=10$ leads to
significantly larger residuals than leaving $n$ free.

\begin {figure*}[t]
\plottwo{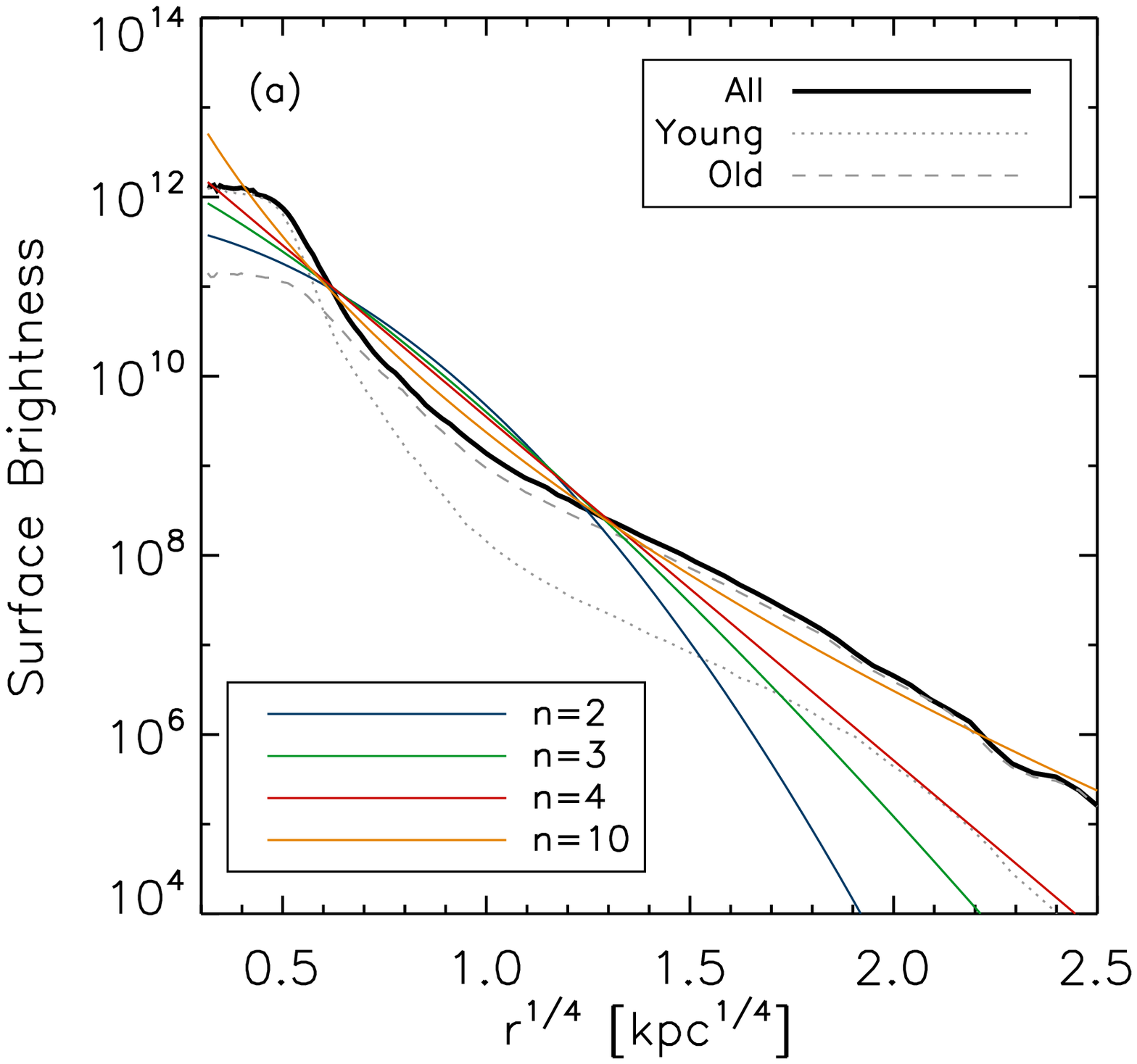}{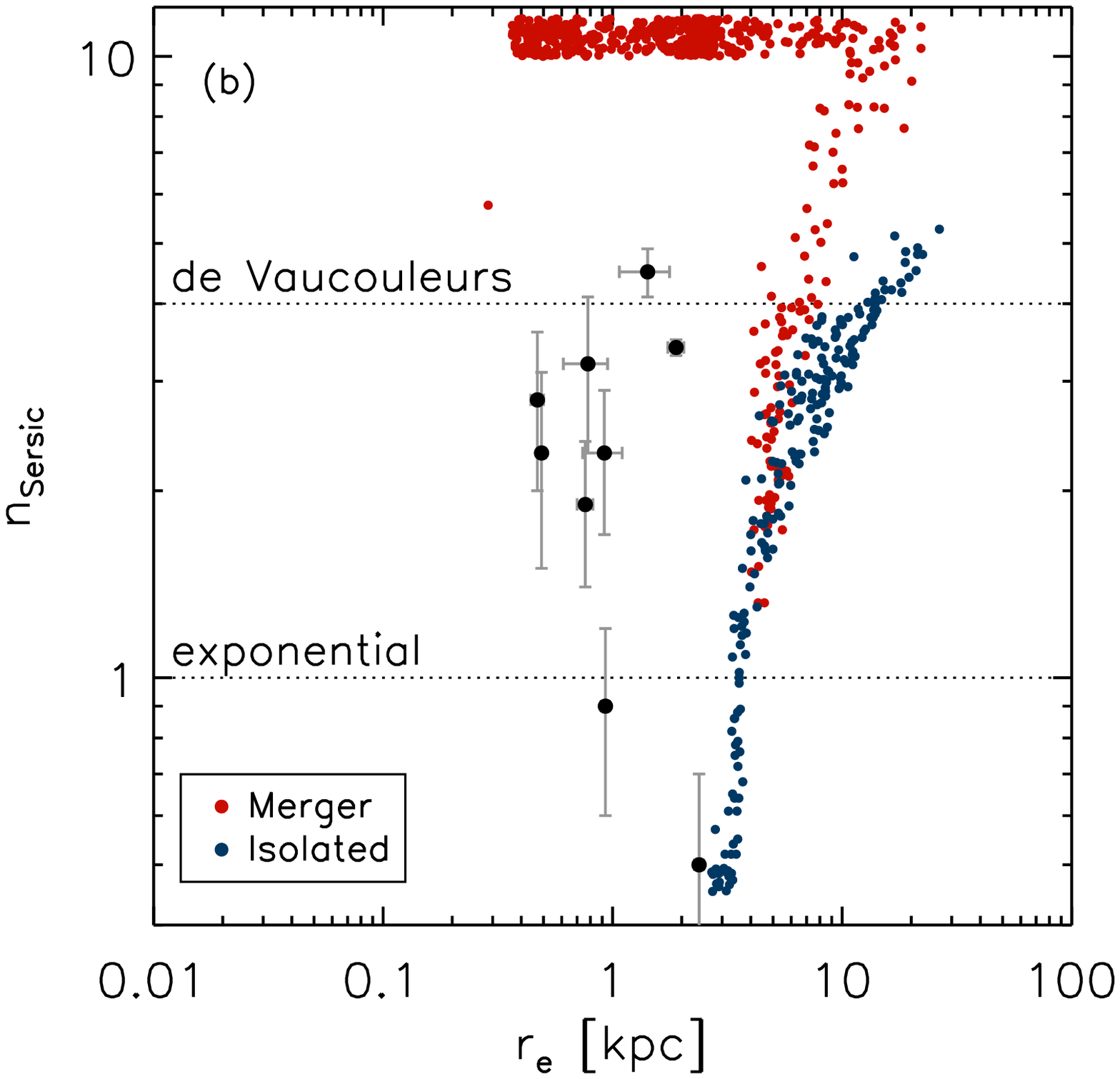}
\caption{(a) Median rest-frame $V$-band surface brightness profile of
a simulated merger remnant ({\it black solid line}), decomposed into a
young (formed during the final burst, {\it dotted gray line}) and old
(formed prior to final coalescence, {\it dashed gray line}) stellar
component.  In color, we overplot Sersic profiles of identical total
luminosity and half-light radius for Sersic indices $n=2$, 3, 4, and
10.  The simulated merger remnant has a different nature than a pure
Sersic profile, but is better approximated by a high value of $n$ than
a low $n$.  (b) Sersic index as function of half-light radius as
derived from GALFIT two-dimensional fitting, for the vD08 sample ({\it
large black symbols}), and binary merger ({\it red dots}) and isolated
disk ({\it blue dots}) simulations of similar mass placed at a similar
redshift ($z = 2.3$).  Neither merger nor disk simulations
simultaneously reproduce the size and profile shape of the observed
quiescent galaxies.
\label{n_re.fig}}
\end {figure*}
Repeating the analysis for random viewing angles of a simulated merger
remnant (Figure\ \ref{sim_stamps.fig}), we obtain strikingly different
results.  For the majority of sightlines, we find that exponential
($n=1$) and even de Vaucouleurs ($n=4$) profiles provide a poor fit to
the mock data.  Their $n_{\rm fix} = 2-4$ residual images (illustrated
in the top 6 rows of Figure\ \ref{sim_stamps.fig}) are characterized
by a central positive peak, surrounded by a negative ring, outside of
which a positive wing is barely visible above the noise.  Our results
remain unchanged when running GALFIT with the sky level left as a free
parameter.  The origin of this characteristic pattern becomes clear
when considering a one-dimensional representation of the surface
brightness, free of the PSF, noise, and pixelization effects of the
mock observations (Figure\ \ref{n_re.fig}a).  Here, the black solid
line indicates the median surface brightness profile of 100 sightlines
uniformly distributed over a sphere.  The dashed and dotted gray lines
decompose the light into that of stars formed prior and during final
coalescence respectively.  As discussed extensively by Hopkins et
al. (2008b; 2009b), the young component produces a central cusp, and
the combination of the young and old component yields a profile that
is poorly characterized by a simple Sersic index.  The colored curves
in Figure\ \ref{n_re.fig}(a) illustrate Sersic profiles with a total
luminosity and half-light radius identical to that of the simulated
merger remnant.  The radially alternating positive-negative-positive
residual pattern is immediately apparent, as is the tendency to fit
high values of $n$.  It is important to note that this tendency is
driven both by the presence of a central cusp (the amplitude of which
depends on the amount of gas present at final coalescence) and by the
presence of the strong wings to the profile (the amplitude of which is
determined by the amount of gas consumed during the progenitor phase).

For a smaller number of sightlines (illustrated by the bottom 3 panels
of Figure\ \ref{sim_stamps.fig}) a flat residual image is obtained
when fitting a Sersic index in the range $2 < n < 5$.  Obscuration of
the central young component by dust is responsible for this
sightline-dependent effect.

We demonstrate our findings with more statistical robustness (more
lines of sight and timesteps, and simulations with different orbital
configurations) in Figure\ \ref{n_re.fig}(b).  Here, we plot the
profile shape (characterized by the best-fit Sersic index) as function
of the half-light radius recovered by GALFIT.  Large black circles are
the vD08 compact galaxies.  They have a median $n = 2.3$, well below
the $n=4$ de Vaucouleurs profile that is characteristic for nearby
spheroids.  van Dokkum et al. (2010) attribute this evolution to minor
merging building up the wings of the high-redshift compact nuggets
over cosmic time.  The Sersic profiles fitted to the simulated merger
remnants (red dots) are significantly cuspier (higher $n$, adopting an
upper bound of $n = 10$).  For the subset (about one fifth) of
sightlines where the central cusp is sufficiently obscured by dust,
the inferred half-light radius is several kpc, inconsistent with the
observed compact systems.

\begin {figure}[h]
\plotone{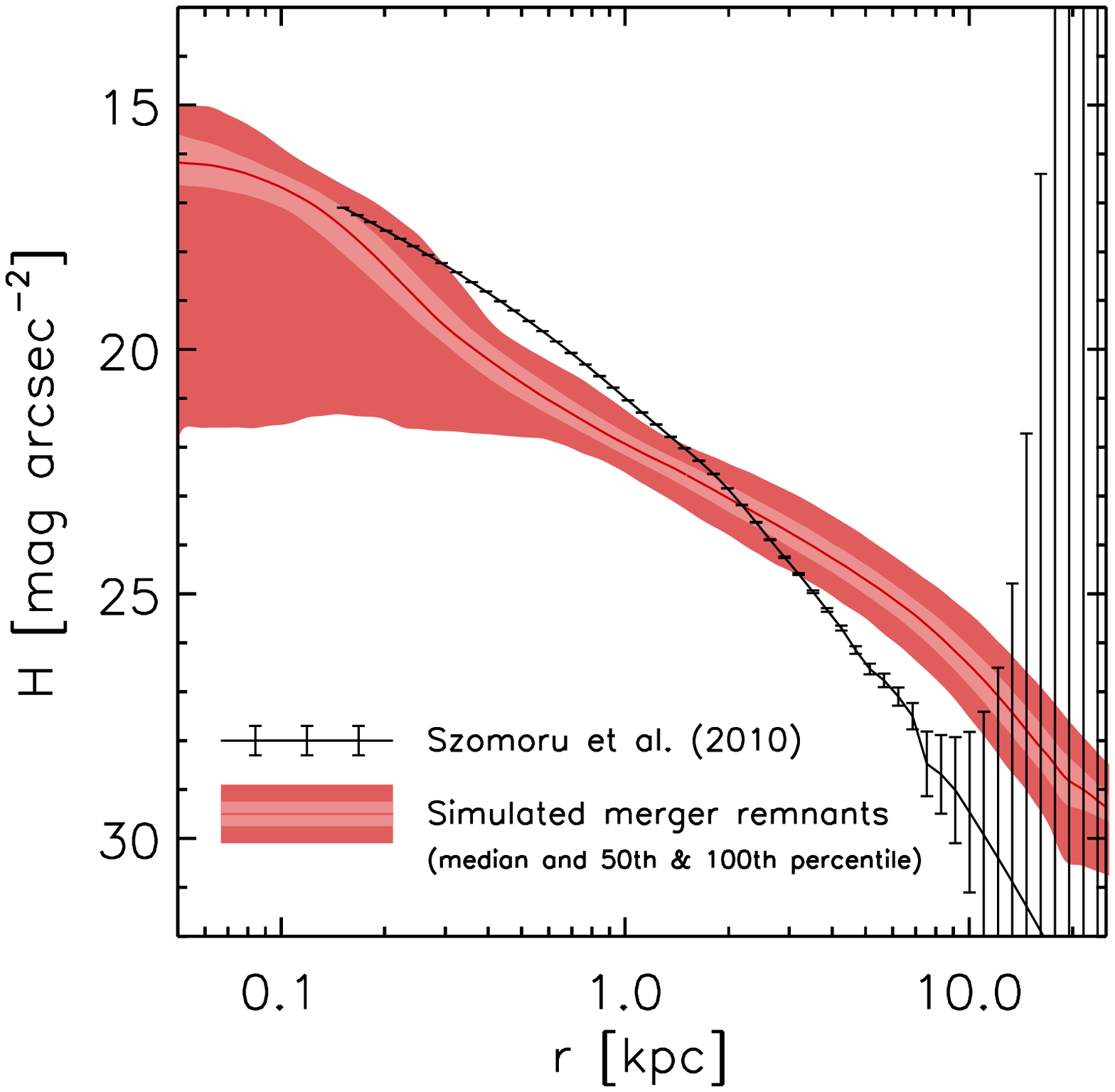}
\caption{Direct comparison of the surface brightness profiles of simulated
merger remnants with the deconvolved profile of a massive, compact
galaxy at $z=1.91$ in the Hubble Ultra Deep Field (Szomoru et
al. 2010).  The profile of the observed quiescent galaxy falls off
more rapidly at large radii than the simulated profiles.
\label{szomoru.fig}}
\end {figure}
We draw a similar conclusion from a direct comparison of the surface
brightness profiles of our simulated merger remnants to that of a
massive compact quiescent galaxy at $z = 1.91$ in the Hubble Ultra
Deep Field (Szomoru et al. 2010, see Figure\ \ref{szomoru.fig}).
Szomoru et al. (2010) derive the deconvolved profile of this galaxy
from very deep Wide Field Camera 3 (WFC3) imaging using a novel
technique that corrects the best-fit Sersic profile with the residual
of the fit to the observed image.  They find that the surface
brightness profile is well approximated by an $n=3.7$ Sersic profile.
Typically, our simulated merger remnants have a higher surface
brightness in the $r > 2$ kpc wings relative to the surface brightness
in the $0.2 < r < 2$ kpc range.

The ability to correctly recover the true half-light radius known from
the simulation depends on the upper bound set on the Sersic index.
For $n_{\rm max} = 10$, we find a systematic overestimate of the size
by $\sim 40$\%, whereas setting $n_{\rm max} = 4$ improves the size
measurement to about $\sim 10$\% in the median, with a scatter in
$\frac{\Delta r_e}{r_e}$ of $\sim 0.3$.  So, whereas the mock images
for the majority of sightlines are best fit by high values of $n$,
these best fits do not properly recover the true half-light radius of
the system.  This emphasizes the fact that an individual Sersic
profile poorly describes its structure.

\begin {figure*}[htb]
\plottwo{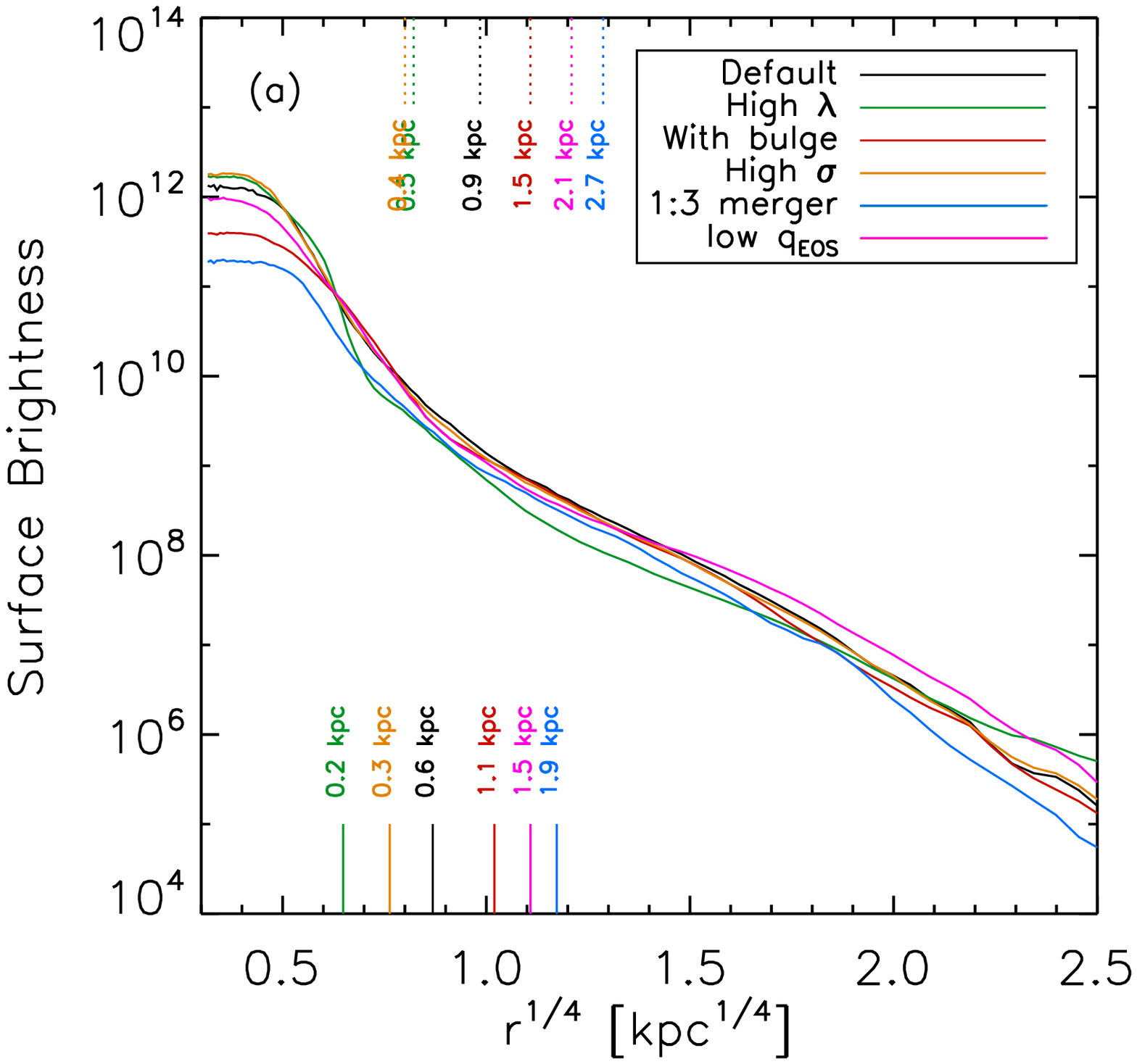}{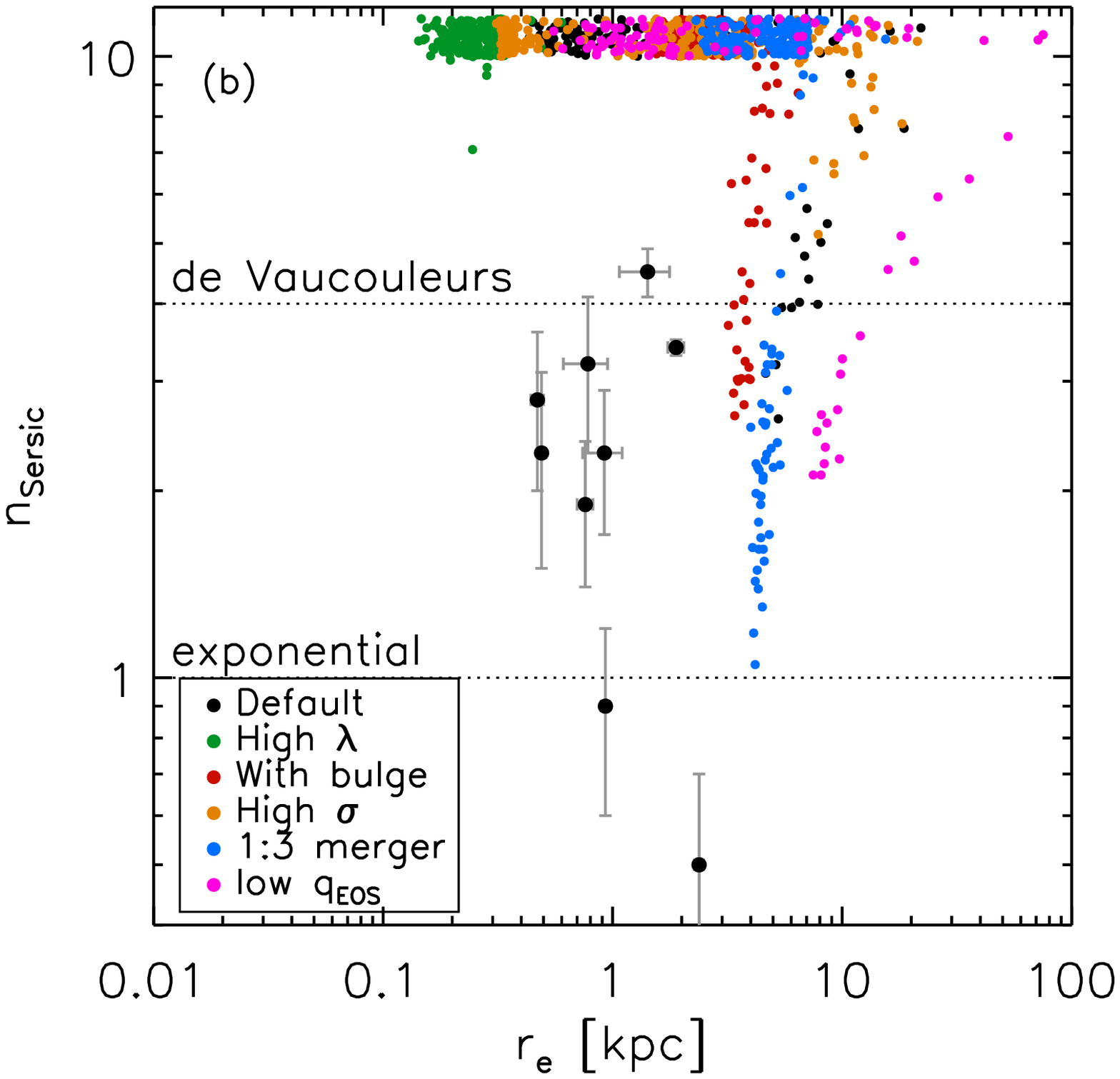}
\caption{Idem as Figure\ \ref{n_re.fig}, but for a series of
simulations where the progenitor properties were varied.  Vertical
lines at the bottom of panel (a) indicate the half-light radii in the
rest-frame $V$-band for which the surface brightness profiles are
plotted.  Dotted vertical lines on the top of panel (a) indicate the
radii containing half of the rest-frame $U$-band light.  When running
GALFIT on mock NIC2 observations of the simulations placed at $z
\sim 2.3$, the inferred Sersic index and size (b) span a different
part of parameter space than covered by the vD08 sample of observed
quiescent galaxies ({\it large black circles}).
\label{progenitor.fig}}
\vspace{-0.1in}
\end {figure*}
We assessed the impact of observational depth by artificially
decreasing the signal-to-noise ratio of our mock observations (in
practice, we reduce the source brightness, and add them to empty
regions of the same NIC2 images).  Running GALFIT with $n_{\rm
max}=10$, this yields an increase in both the systematic size
overestimate and the scatter in $\frac{\Delta r_e}{r_e}$ over a range
of viewing angles.  Adopting $n_{\rm max}=4$, we find that, in the
median, the recovered size is consistent with the true size over the
full range of input source magnitudes $21 < {\rm mag_{F160W}} < 24$.
However, the scatter in $\frac{\Delta r_e}{r_e}$ does increase by a
factor 2-3 as we compare sources with ${\rm mag_{F160W}} = 24$ to
sources of identical profile shape with ${\rm mag_{F160W}} = 21$.  We
caution that different biases may be at play in the size measurement
of observed compact quiescent galaxies such as the vD08 sample.  Their
intrinsic profile shape (i.e., noise-free and at infinite resolution)
is unknown, but we did demonstrate it differs from that of our sample
of simulated merger remnants as they translate into different (mock)
observed properties.  As an independent test, we also inserted pure
Sersic profiles of $\sim 1$ kpc size, with $1 \leq n \leq 4$ and a
range of luminosities into empty regions of the NICMOS images.
Running GALFIT on them, we find qualitatively similar results
regarding potential biases in size measurements.  The scatter in the
recovered size increases with decreasing signal-to-noise ratio.
Whereas profiles of similar magnitude as the galaxies from vD08 ($21 <
{\rm mag}_{F160W} < 22$) were recovered correctly, systematic
overestimates of the size by a factor 1.06 (1.2) in the median and 1.1
(2.2) in the mean are found when decreasing the brightness of
non-exponential ($2 \leq n \leq 4$) sources by 1 (2) magnitudes.  In
other words, perhaps counter-intuitively, size measurements from 2D
surface brightness fitting on low signal-to-noise data may be biased
towards size estimates that are too large, rather than too small.

Computing the mock observations of our simulated merger remnants using
the LOS and SUNRISE radiative transfer codes yields qualitatively
similar results.  Namely, that the simulated surface brightness
profiles tend to be cuspier, and in general occupy a different locus
in ($n$, $r_e$)-space than the observed compact galaxies at $z \sim
2$.  Moreover, we confirm that this difference is notable even given
the limitations in depth and resolution of present-day observations.
In detail, we note that, for a given simulation, the LOS code produces
cuspier profiles than SUNRISE, yielding significant residuals after
subtracting a $n=10$ fit (e.g., rows 1-3 of Figure\
\ref{sim_stamps.fig}) whereas these are mostly flat when running
GALFIT on SUNRISE images (e.g., rows 4-6 of Figure\
\ref{sim_stamps.fig}).

Finally, we carried out a similar procedure on mock images of
simulations where we let disk galaxies evolve in isolation.  The
initial disks are designed to have an exponential mass profile.  We
checked that GALFIT correctly recovers their $n=1$ profile shape from
mock images of the first simulation snapshot (values of $0.6 < n < 1$
are recovered for some sightlines due to dust obscuration in the
central region).  Eventually, since gas is consumed by star formation
and we did not allow for gas replenishment, the simulated disk
galaxies also reach low specific star formation rates.  By the time
this quiescent phase is reached, the disks have evolved in such a way
that we find a range of Sersic indices similar to that of the vD08
sample (see Figure\ \ref{n_re.fig}b).  However, in all cases, their
recovered size is a factor of several too large to reproduce the
observed quiescent galaxies.

\subsection {Dependence on progenitor properties}
\label{progenitor.sec}

We now consider a few variations on the formation scenario for compact
quiescent galaxies, with an emphasis on how they alter the consistency
with observational constraints on size and profile shape.  First, the
apparent discrepancy in profile shape of the merger remnants may stem
from inadequate assumptions on the structure of the progenitors
from which they formed.  Figure\ \ref{progenitor.fig} presents a
qualititative, but by no means complete, exploration of the vast
parameter space in which progenitor properties can vary.  First, we
ran a simulation where the spin parameter of the dark matter halos was
increased from $\lambda = 0.033$ to $\lambda = 0.1$ ({\it green}),
with otherwise default settings.  The higher spin parameter has a
stabilizing effect on the gas disk, lowering the star formation rate
during the progenitor phase, and increasing the gas fraction at the
time of coalescence ($f_{\rm gas} = 0.64$ compared to $f_{\rm gas} =
0.42$ for our fiducial model).  The result is an even more
concentrated remnant.  Less stars formed prior to the merger, but they
were redistributed in such a manner that it is mainly at intermediate
radii (around 1 kpc) that the surface brightness profile is devoid of
light with respect to our default simulation.  This is where the
residual to a Sersic fit with $2 < n < 4$ was already negative (see
Figure\ \ref{n_re.fig}a), causing the best fit to run into the upper
bound set on $n$ again.
Second, we investigate a simulation with the same total baryonic mass,
but with progenitor galaxies that are built up of a stellar bulge in
addition to a gas-rich disk, each contributing half of the baryonic
mass ({\it red}).  Secular formation processes to grow such a bulge by
coalescence of clumps in gas-rich disks (i.e., without invoking
additional merging) have recently been proposed by Genzel et
al. (2008), Elmegreen, Bournaud \& Elmegreen (2008, and Dekel et
al. (2009b).  We find that merging the bulge$+$disk systems leads to a
compact ($r_{e, V{\rm rest}} = 1$ kpc) remnant with significantly
smoother central profile, but the strong wings remain in place.
Analyzing its mock observations with GALFIT yields a locus in ($n$,
$r_e$)-space that does not overlap with the vD08 sample (see Figure\
\ref{progenitor.fig}b).

The same conclusion can be drawn from a third and fourth variation
where we increase the disk dispersion twofold ({\it orange}) and
adjust the merger mass ratio from equal mass to 1:3 ({\it blue}) respectively.
The latter variations are inspired by the small
$\frac{v}{\sigma}=2-4$ ratios of star-forming disk galaxies at $z
\sim 2$ observed by the SINS survey (F\"{o}rster Schreiber et
al. 2009), and by the paucity of equal-mass mergers compared to
mergers of mass ratio 1:3 (Naab et al. 2007; Guo \& White 2008;
Hopkins et al. 2010b).  We do not consider smaller mass ratios as they
would not quench the star formation and produce a quiescent spheroid.

Finally, we considered the effect of adopting a lower pressurization
of the ISM ($q_{EOS} = 0.35$ instead of $q_{EOS}=1$ in the
nomenclature of Springel, Di Matteo \& Hernquist 2005).  This leads to
enhanced star formation in the outer regions during the early stages
of the merging process, resulting in more pronounced wings in the
remnant.  This is in qualitative agreement with Bournaud et al. (2010)
who find that the remnants of mergers of clumpy disks are cuspier
(i.e., have higher Sersic indices and more material at large radii)
than those produced by merging disks in which the effective ISM is not
allowed to fragment so completely.

Overall, the considered variations in progenitor properties have only
a modest impact on the structure of the merger remnants: they occupy a
similar region in the size-mass diagram, and have $M/L$ ratio
gradients that translate into similar color gradients (i.e., red
cores; the dashed and solid lines in Figure\ \ref{progenitor.fig}a
indicate the rest-frame $U$ and $V$ half-light radii respectively).
Moreover, whereas the amplitude of the central cusp shows a dependence
on the specific variations in progenitor structure explored here, the
extended outskirts of the light distribution are present in each of
the above model variations.  As discussed in Section\
\ref{summary.sec}, a wider exploration of parameter space, involving
physics leading to relative changes in the star formation efficiency
between low- and high-density environments, may be appropriate to
structurally alter the outer regions of the merger remnants.

\subsection {The impact of subsequent merging}
\label{subsequent.sec}

\begin{figure}[hbtp]
\plotone{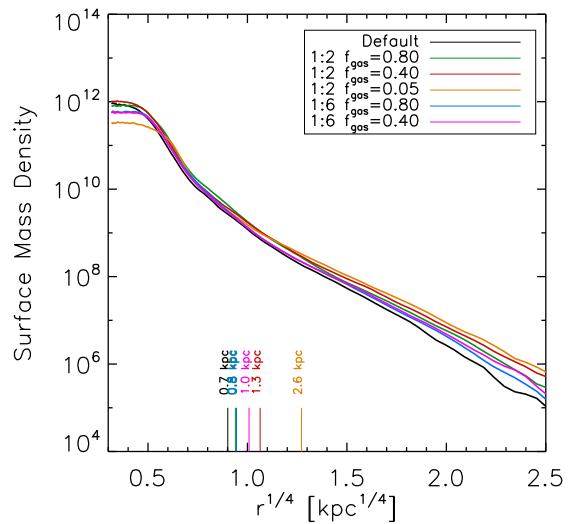}
\caption{Surface mass density profile of a gas-rich merger remnant
({\it black}), and the products after re-merging it with star-forming
galaxies of varying mass and initial gas fraction ({\it colored
curves}).  Half-mass radii are indicated by the vertical lines.
Subsequent merging of a compact gas-rich merger remnant leads to a
further build-up of its outer regions while the central dense cusp is
little affected.  This leads to an increase of the size, an effect
that is most pronounced when the amount of dissipation is minimal.
\label{subsequent.fig}}
\end{figure}
In this section, we address whether subsequent merging can alter the
structure of the simulated merger remnants in such a way that their
profile shape shows a closer resemblance to that of observed quiescent
galaxies.  To this extent, we run a set of re-merger simulations in
which the remnant of the initial gas-rich merger collides with a
star-forming galaxy of lower mass.  We consider mass ratios 1:6 to
1:2, and initial gas fractions of 80\%, 40\% and 5\% for the
star-forming galaxies with which the merger remnant interacts.  The
resulting surface mass densities are presented in Figure\
\ref{subsequent.fig}.  A more extended analysis, involving merger
rates as function of mass, mass ratio, redshift and
(redshift-dependent) gas fraction is required to properly model
spheroid evolution (Khochfar \& Silk 2006b; Naab et al. 2009; Hopkins
et al. 2010a), but for our present purpose it is sufficient to
highlight two characteristic features notable in our re-merger
simulations.

First, the re-merger is consistently larger than the original merger
remnant.  The relation between the half-mass radius before and after
depends on the nature (mass ratio, gas fraction, ...) of the
re-merger.  As expected, the size growth is larger for the 1:2 than
for the 1:6 mass ratios.  At a given mass ratio, the growth is more
pronounced when the amount of dissipation is minimized (i.e., dry
merging is more efficient at expanding galaxies).

Second, the re-merger remnant is not homologous to the original
gas-rich merger remnant (see also Hopkins et al. 2009c; Hopkins \&
Hernquist 2010).  Typically, we see the build-up of an outer wing
while the dense central cusp is only little affected.  This effect is
more pronounced as we consider re-mergers with a lower dissipational
fraction.  The net effect is that, if anything, higher values of $n$
are preferred in fitting the products of subsequent merging.  In
addition, their size at a given mass grows too big to be consistent
with the observed spheroids at $z \sim 2$.  While subsequent (minor)
mergers might well drive the structural evolution of spheroids from $z
\sim 2$ to the present day (Naab et al. 2009; van Dokkum et al. 2010;
Hopkins et al. 2010a; Carrasco et al. 2010), they are not a viable
mechanism to produce the $z \sim 2$ compact galaxies in the first
place.  Even if such an external process would yield sizes and profile
shapes that matched the observations, it would not have offered a
satisfactory solution, since it is unlikely that every quiescent
galaxy observed at $z \sim 2$ experienced a second interaction after
the initial quenching event.

\subsection {Stellar mass loss}
\label{massloss.sec}

In light of the last comment, it is interesting to consider internal,
rather than external, processes that may alter the light distribution.
One such process is stellar mass loss.  The simulations presented in
this paper were run with GADGET-2's default settings that account for
a small amount of mass loss.  Ten percent of the gas mass converted
into stars is instantaneously returned to the interstellar medium
(ISM, Springel \& Hernquist 2003).  With it, energy is injected in the
surrounding ISM to model the effect of short-lived stars that explode
as supernovae.  For currently favored stellar initial mass functions
(IMFs) such as a Kroupa (2001) or Chabrier (2003) IMF, however,
eventually as much as half of the gas mass initially converted into
stars is returned to the ISM.  Initially, the mass loss proceeds fast
and in an explosive fashion, reaching 10\% in 7 Myr and 25\% in 60
Myr.  At later times, AGB stars drive more gradual winds, increasing
the released fraction to $\sim 40$\% after a gigayear (Bruzual \&
Charlot 2003).  In principle, three things can happen to the mass lost
due to winds: either it escapes the galaxy's potential entirely, or it
settles within the galaxy and stays there as gas, or it is consumed by
star formation again.  In addition, the redistribution of baryonic
mass due to mass loss can potentially change (lower) the central
potential, causing the distribution of stars to adjust (expand)
accordingly.  Fan et al. (2008) propose such a scenario of adiabatic
expansion due to mass loss driven by AGN and/or stellar feedback as a
mechanism for spheroid growth between $z \sim 2$ and $z \sim 0$.
Others (Bezanson et al. 2009; van Dokkum et al. 2010; Hopkins et
al. 2010a) have objected to this process as a driver of structural
evolution since $z \sim 2$, amongst other reasons because most of the
mass loss is expected to take place within 500 Myr after the stars are
formed, i.e., prior to the phase when the galaxies are observed as
being quiescent.  The same timescale argument works in favor of our
present discussion.  If mass loss modifies the surface brightness
profile substantially, it will have happened by the time we observe
the galaxies as being quiescent.

The precise effect of such extensive mass loss is however not trivial
to predict, as the energetics, the amount of mass loss, and the
location of the stars when they lose mass all depend on time.
Moreover, the timescales of stellar evolution (i.e., mass loss), star
formation history of the system, and merger dynamics all overlap.  It
will be interesting to investigate in future numerical work how a
time-dependent treatment of extensive stellar mass loss and its
feedback affects the outcome of simulations as the ones studied here.
Since this is beyond the scope of this paper\footnote[3]{We did test
that, ignoring any time dependence of mass loss, increasing the
fraction of material instantaneously returned to the ISM from 10\% to
25\% barely changes the surface brightness profile of the remnant.}, we
now limit ourselves to simple toy models addressing the impact of,
e.g., feedback on how the simulated quiescent galaxies would be
observed.

\subsection {Feedback, radiative transfer, and preventing star formation at early times}
\label{implementation.sec}

\begin {figure*}[t]
\plottwo{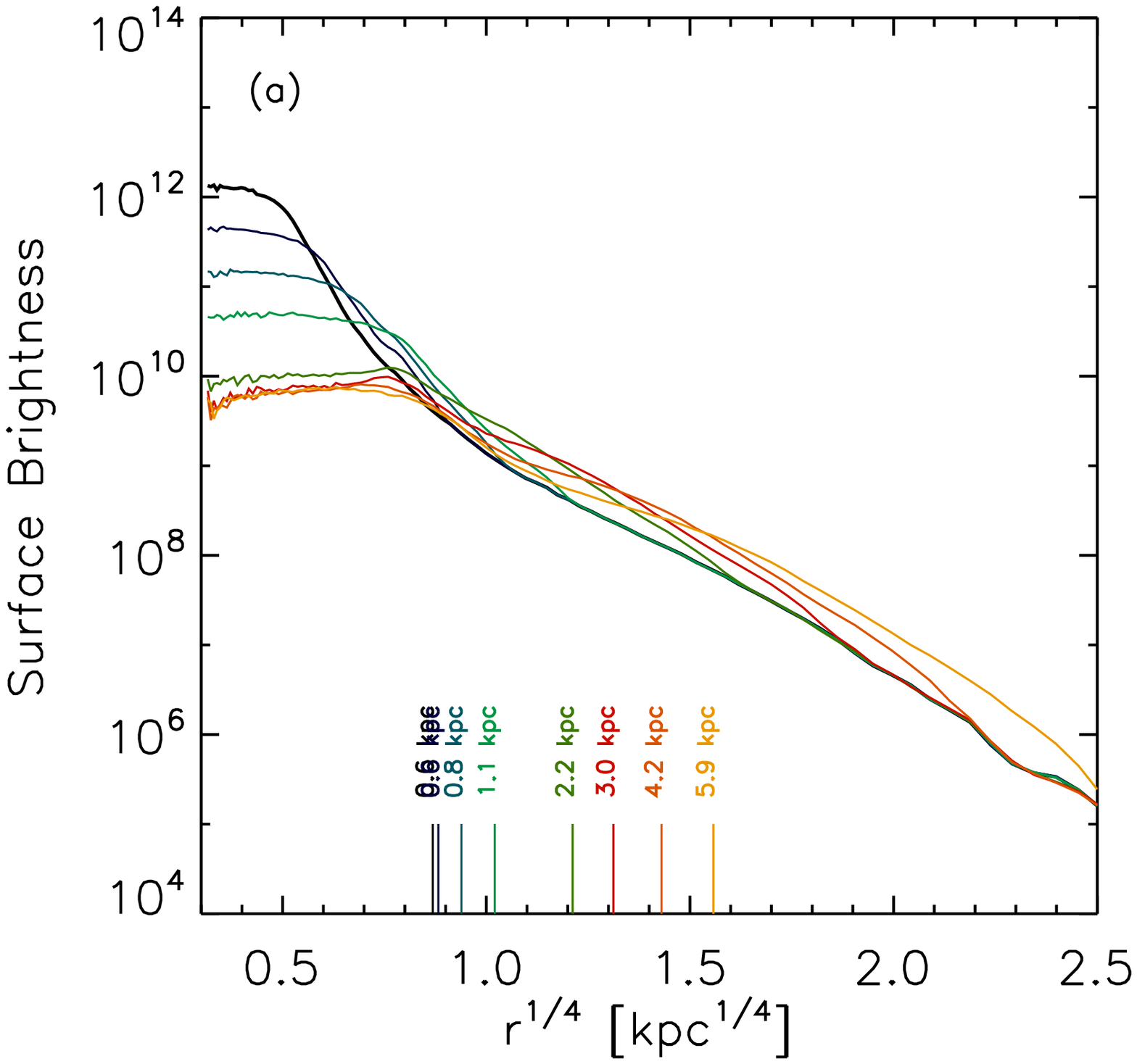}{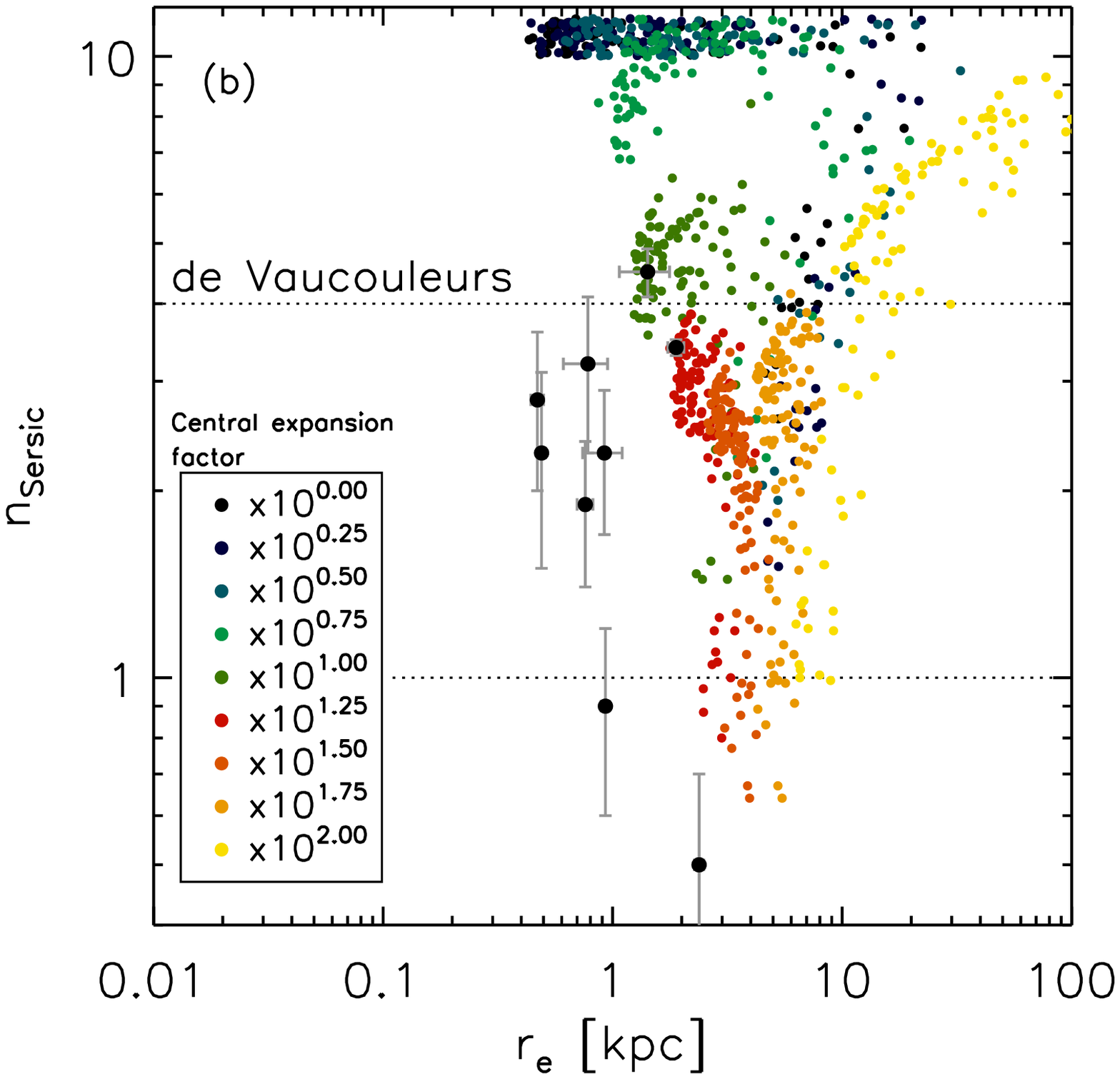}
\plottwo{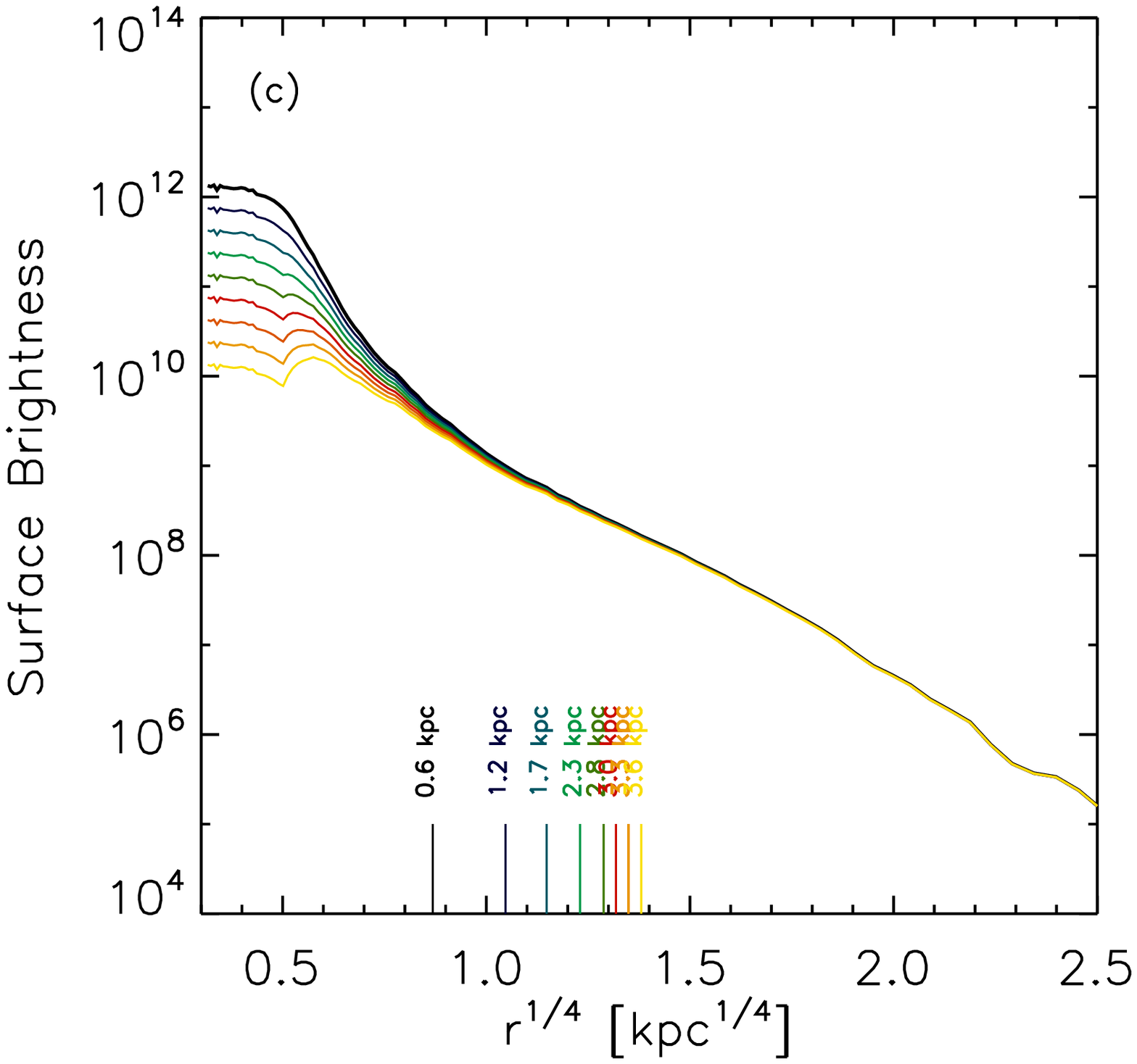}{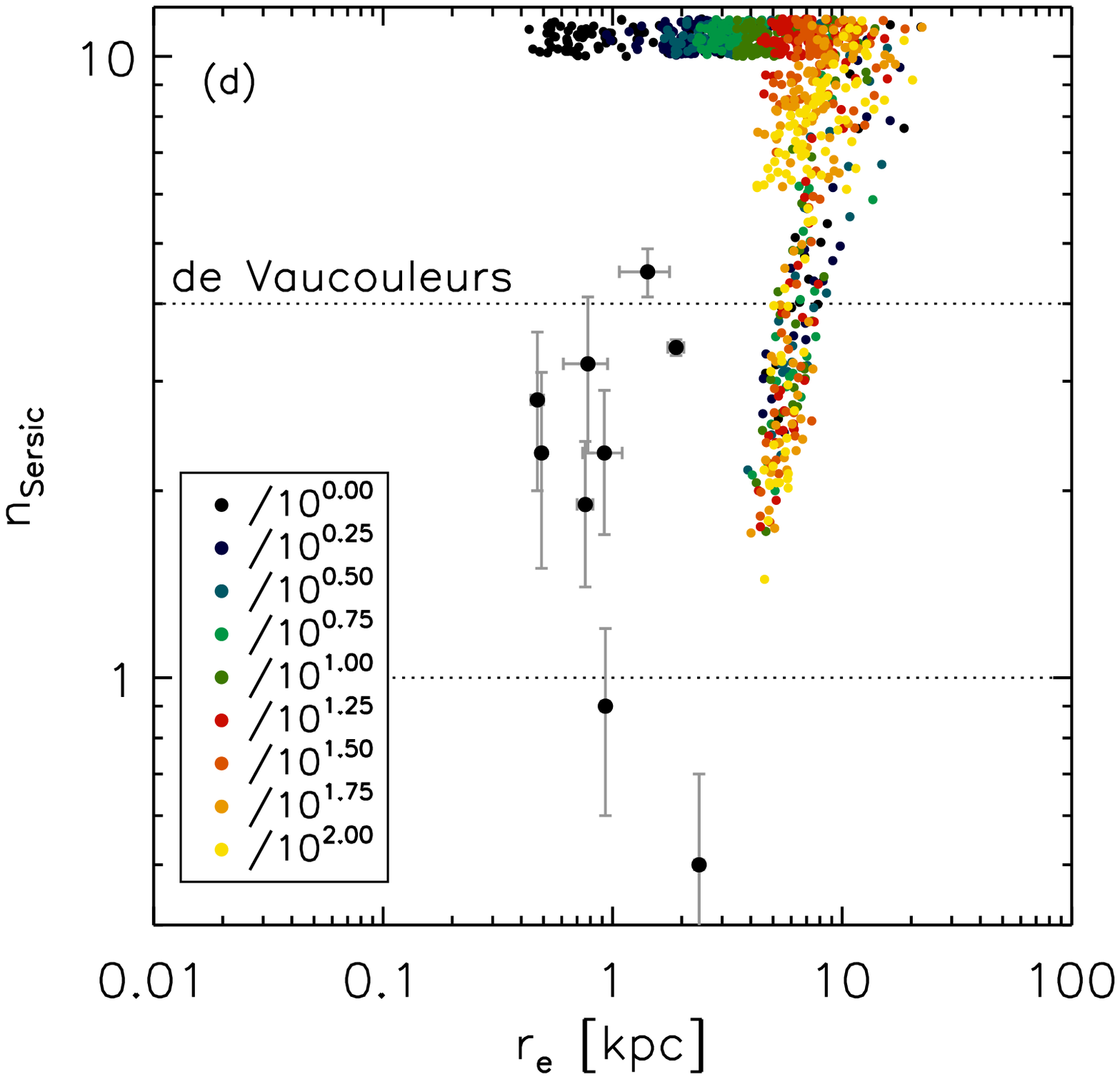}
\plottwo{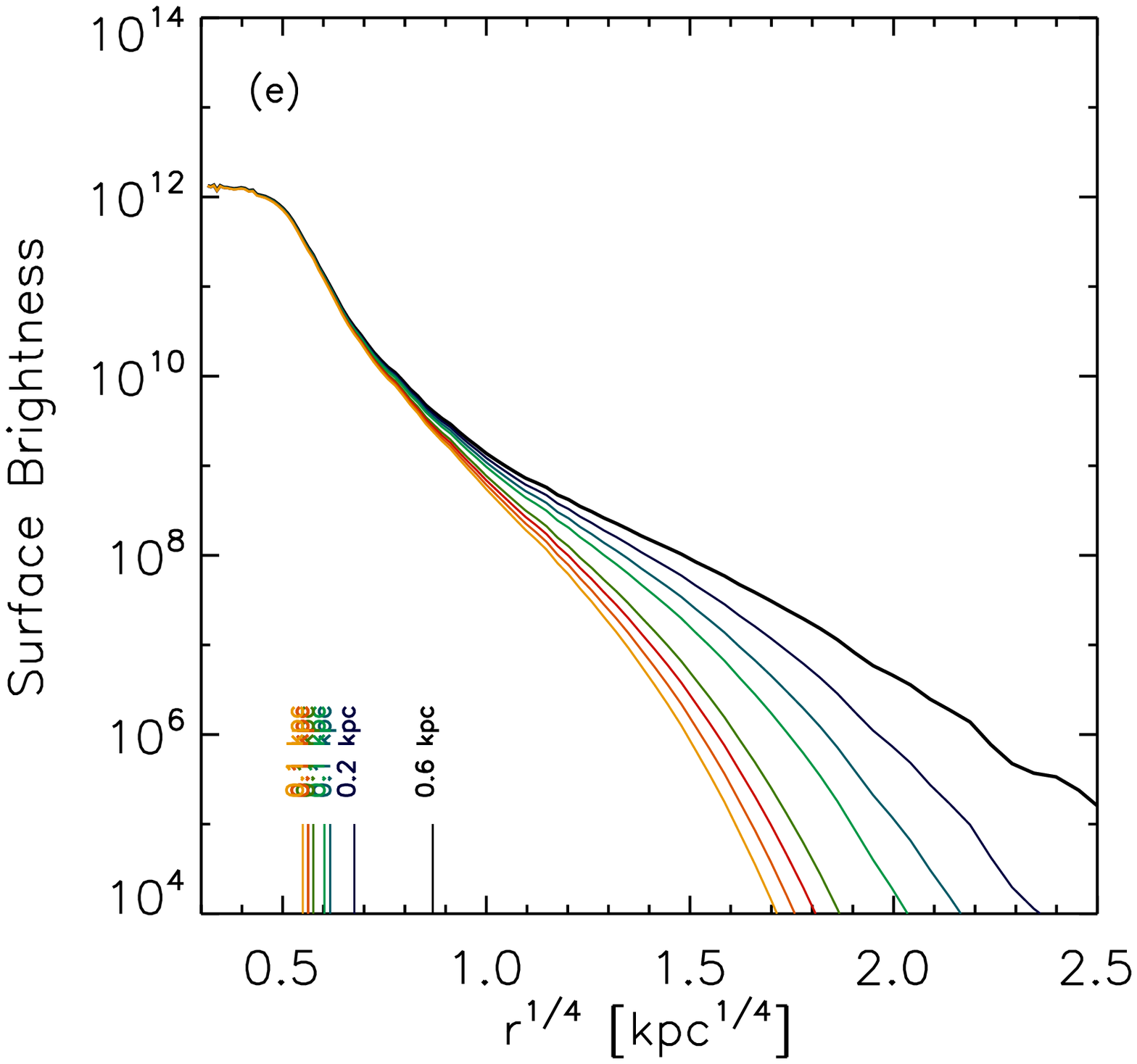}{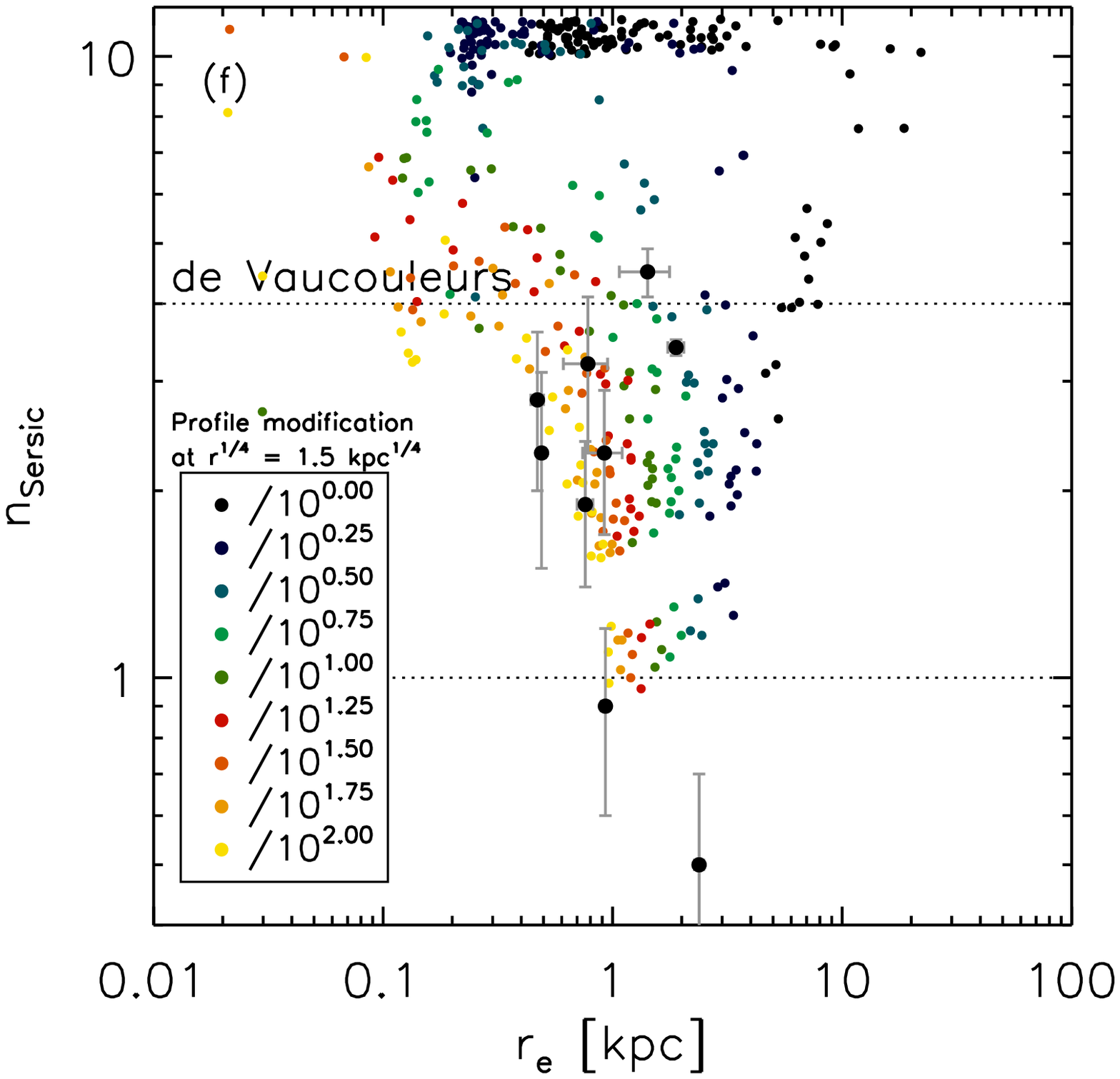}
\caption{\small Idem as Figure\ \ref{n_re.fig}, but for three toy models
where we apply a simple transformation to a snapshot from a fiducial
simulated merger remnant: (a-b) an expansion of the central region,
(c-d) radially decreasing extinction, (e-f) radially increasing
extinction.  The high Sersic indices preferred in fitting the
two-dimensional surface brightness profiles are driven by both the
central cusp and the amount of light emerging from the extended wings
of the profile.  Only adjusting the central profile shape (a-d) is
insufficient to match the vD08 galaxies in ($n$, $r_e$)-space.  A
mechanism to reduce the amount of stars at large radii (typically
formed in the progenitors long before final coalescence), or the
amount of light emerging from them seems required.
\label{feedback_rt.fig}}
\end {figure*}
In these toy models, we apply a simple transformation to a simulation
snapshot of a merger remnant: an expansion by a fixed factor of the
central component where the young stars dominate (Figure\
\ref{feedback_rt.fig}a), and additional extinction that decreases
(Figure\ \ref{feedback_rt.fig}c) or increases (Figure\
\ref{feedback_rt.fig}e) with radius.  The first toy model may be
thought of as an ad hoc implementation of feedback, whereas the latter
two explore dramatic differences in the effect of radiative transfer.
The corresponding diagrams of recovered Sersic index versus recovered
size when feeding mock NIC2 observations of the toy models to GALFIT
are presented in the right-hand panels.  The color-coding traces the
expansion or extinction factor, where the legend quotes the extinction
at $r^{1/4} = 0.5$ kpc$^{1/4}$ and $r^{1/4} = 1.5$ kpc$^{1/4}$ for the
radially decreasing and increasing extinction respectively.

Figure\ \ref{feedback_rt.fig}(b) shows that, if a physical mechanism
that is currently not or improperly modeled is capable of puffing up
the central stellar cusp by over an order of magnitude, the best-fit
Sersic indices enter the regime of those of the observed compact
galaxies.  However, the same transformation would lead to a violation
of the observational constraints on size by a factor of several.

Likewise, larger sizes are inferred from toy model images where the
central region is heavily obscured (Figure\ \ref{feedback_rt.fig}d).
Moreover, as we already noted in the case of merging bulge$+$disk
progenitors (\S\ref{progenitor.sec}), a reduction of the central cusp
without any change to the outer wings does not necessarily lower the
Sersic index inferred from the mock observations of limited depth and
resolution.

A better correspondence with the locus of the vD08 sample in ($n$,
$r_e$)-space is obtained when we lower the amplitude of the profile
wings (Figure\ \ref{feedback_rt.fig}e).  The plotted transformations
reduce the true half-light radius (vertical lines in Figure\
\ref{feedback_rt.fig}e) by a factor 2 to 7, but the sizes inferred by
GALFIT are of order 1 kpc.  Although these toy models have observed
properties similar to the vD08 galaxies, the required transformation
factors are large (an order of magnitude or more) and it is hard to
think of a radiative transfer effect related to the dust distribution
or intrinsic stellar population properties in the wings that can
account for such a deviation from our fiducial model.

There is another way of interpreting the last toy model.  The
outskirts are built up of old stars that formed in the progenitor
disks and were redistributed by violent relaxation during the merger.
Consequently, limiting the amount of star formation at early times may
lower the amplitude of the wings.  In the binary merger simulations
considered here, this is hard to accomplish since all gas is already
present at the start of the simulation and the compact gas-rich disks
are sensitive to gravitational instabilities giving rise to star
formation.  This is the case even for the fully-pressurized multiphase
ISM model adopted throughout this paper ($q_{\rm EOS} = 1$).  A way to
circumvent this problem would be to delay the supply of gas or spread
it more continuously over time.  Indeed, recent cosmological
simulations suggest that galaxies during the first 3 billion years
after the Big Bang were not closed systems, and supply of gas through
cold filaments and streams was a generic feature of their evolution
(Keres et al. 2005; Dekel \& Birnboim 2006; Ocvirk et al. 2008; Dekel
et al. 2009a).  As such, reproducing not only the stellar population
properties and size-mass relation, but also the profile shape of
high-redshift quiescent galaxies may require simulations in a
cosmological setting.  This is numerically very challenging as a high
spatial and timestep resolution (see \S\ref{sim_main.sec}) remain
essential to model the detailed structure of the merger remnants, as
well as the accretion processes onto the supermassive black hole(s)
that contribute to the quenching of their star formation.
Consequently, a resimulation technique (see, e.g., Tormen, Bouchet \&
White 1996) may be appropriate, as applied for example by Naab et
al. (2007, 2009) in the context of spheroid formation.

\vspace{0.1in}

\begin {figure*}[t]
\plotone{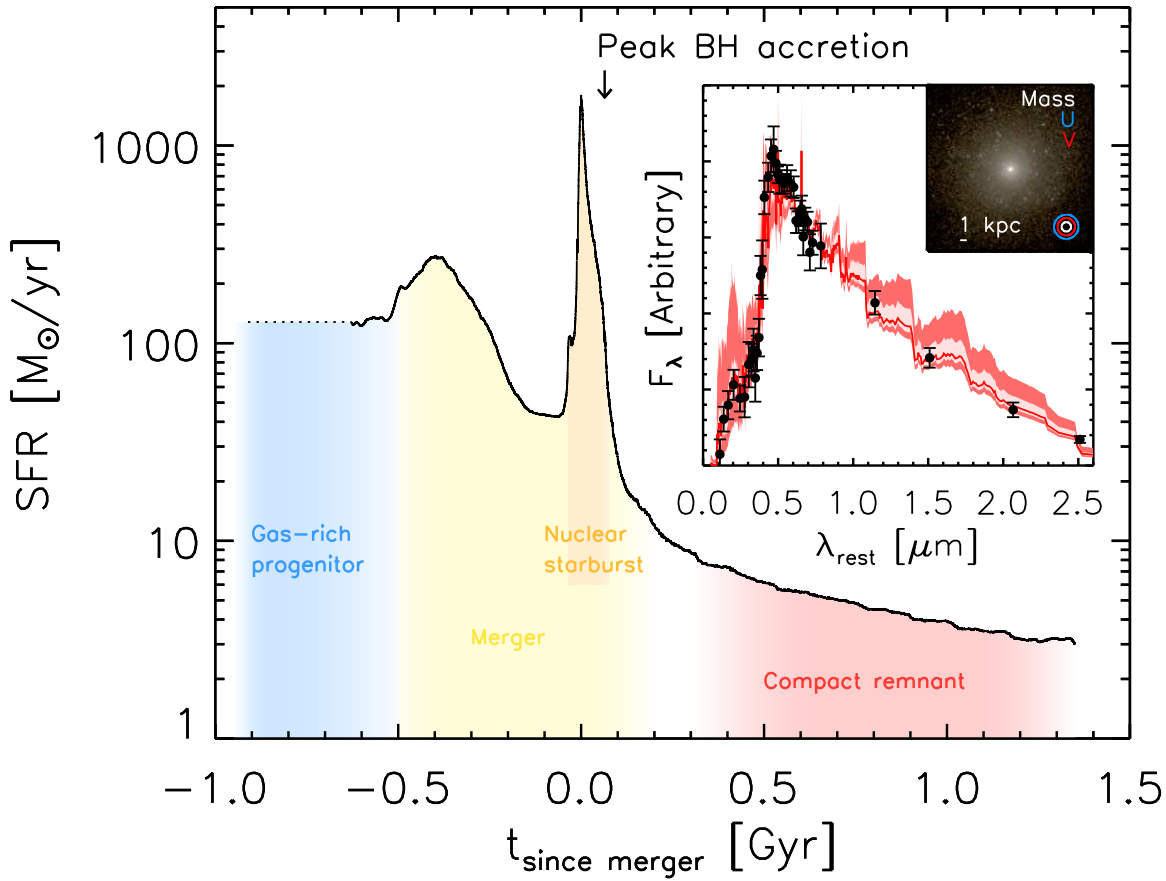}
\caption{Star formation history of a gas-rich merger simulation
producing a compact remnant.  The spectral energy distribution (SED)
of simulated remnants ({\it red in inset panel}) provides a good match
to the mean SED of the observed compact quiescent galaxies from vD08.
White, blue and red circles in the log-scaled postage stamp show the
circularized region containing half the total mass, $U$- and $V$-band
light respectively.  A mass-to-light ratio gradient is present in the
simulated remnant, which typically results in larger half-light radii
at shorter wavelengths.
\label{summary.fig}}
\vspace{-0.1in}
\end{figure*}

\section {Discussion and summary}
\label{summary.sec}

The existence of ultra-compact, massive ($\sim 10^{11}\ M_{\sun}$)
quiescent galaxies already in place merely 3 billion years after the
Big Bang has been one of the great surprises of galaxy evolution
studies in the last few years.  Their discovery has triggered
questions regarding their origin and fate, as well as the reliability
with which their properties such as mass and size are characterized.

In this paper, we address the formation and nature of such systems in
the context of (gas-rich) merger scenarios.  As such, this study is
part of a larger effort that couples hydrodynamic simulations with
radiative transfer in order to investigate how different types of
observed high-redshift galaxies may tie together as stages of one
evolutionary sequence.  Figure\ \ref{summary.fig} illustrates a
typical star formation history of a gas-rich merger simulation.
Gas-rich star-forming galaxies ({\it blue}) are abundantly present at
high redshift (Erb et al. 2006; Tacconi et al. 2010).  When in the
process of merging ({\it yellow}), the morphology of such systems may
(but will not always) appear disturbed.  The timescales of a
morphological merger signature depend on mass ratio (Lotz et
al. 2010a) and gas fraction (Lotz et al. 2010b) of the merger.
Narayanan et al. (2010) describe how during a relatively short ($< 50$
Myr) period at final coalescence ({\it orange}) a peak in emission at
sub-millimeter wavelengths is reached.  Continuing over a somewhat
longer ($\sim 100$ Myr) timespan, the system may be identified as a
so-called dust-obscured galaxy (DOG), characterized by its large 24
$\mu$m to optical flux ratio, although not every DOG, and particularly
those with low 24 $\mu$m luminosity, is a gas-rich merger (Narayanan
et al. 2009).  Shortly after the peak in star formation rate (at
$t_{since} = 64$ Myr for the particular simulation shown), the
accretion rate onto the central supermassive black hole peaks.  The
role of merging in galaxy - black hole coevolution has been discussed
extensively by Hopkins et al. (2006, 2008a).  Finally, Wuyts et al. (2009b)
investigated the role major mergers may have played in producing the
above mentioned compact systems, in terms of their number and mass
densities as well as their integrated stellar population properties.
Iterating on the latter aspect, the inset panel in Figure\
\ref{summary.fig} contrasts the distribution of rest-frame UV to
near-infrared SEDs of merger remnants computed from our simulations
({\it median, 50th and 100th percentiles are displayed in red}) to the
mean rest-frame SED of quiescent galaxies from the vD08 sample (Muzzin
et al. 2009b, {\it black data points, with error bars representing the
error on the mean}).  The model and observations show an excellent
agreement over the full wavelength range probed, boosting confidence
that the modeled stellar populations reflect reality at least in an
integrated sense.

In this paper, we focussed on the structure and resolved stellar
populations of merger remnants.  We used the sample of massive
quiescent galaxies from vD08 as reference sample.  Their near-infrared
spectra are characterized by a Balmer/4000\AA\ break (Kriek et
al. 2006).  Deep NIC2 observations (vD08) have revealed their compact
nature at the highest resolution currently available.  Analyzing a
suite of binary merger simulations of varying mass, gas fraction,
progenitor scaling, and orbital configuration, we confirm the idea
originally proposed by Khochfar \& Silk (2006a) that major mergers can
explain their location in the size-mass diagram provided they are
gas-rich.  Merging galaxies scaled to represent high-redshift
star-forming disks, we find that systems of $\sim 10^{11}\ M_{\sun}$
with half-mass radius $\sim 1$ kpc can be formed when the gas fraction
by the time of final coalescence is about $\sim 40$\%.  Observational
evidence for gas fractions of this magnitude has been accumulating in
recent years (e.g., Tacconi et al. 2010).  The corresponding velocity
dispersions of these simulated massive compact galaxies are of order
300 - 400 km s$^{-1}$.  They show considerable rotation ($v_{\rm
maj}/\sigma$ of up to unity) compared to the majority of lower
redshift early-type galaxies, a result that has yet to be confirmed
observationally.

Running radiative transfer on the output of our SPH simulations, we
find that the merger remnants have a radially dependent mass-to-light
ratio.  Typically, when observing the remnant 500 Myr to 1 Gyr after
the peak in star formation rate, the half-mass radius is a factor
$\sim 2$ smaller than the rest-frame $V$-band half-light radius.  This
implies that the high effective densities inferred from NICMOS and WFC3
observations of high-redshift quiescent galaxies may in fact only be
lower limits.  In the rest-frame $U$-band, the typical light-to-mass
size ratio increases to a factor 3-4.  The ratio $r_{e, {\rm light}} /
r_{e, {\rm mass}}$ shows a significant sightline-dependence, with a
tail to high values ($\gtrsim 5$).  Our conclusions are robust to the
choice of radiative transfer code, input stellar population synthesis
models, dust attenuation law, and whether or not the progenitor disks
had stellar population gradients.  The $M/L$ ratio gradient is
expressed as a color gradient that correlates with the integrated
color of the system (redder galaxies hosting a more pronounced red
core).  We caution that the presence of a red core cannot
unambiguously be interpreted as evidence for inside-out growth.  In
fact, in the simulations studied in this paper, the central stellar
population formed last, rather than first.  The effect of the negative
age gradient is compensated by the fact that the stars inside the
half-mass radius are more metal-rich, and suffer more extinction.

Finally, we perform two-dimensional parametric fitting with the GALFIT
code (Peng et al. 2002) on real and mock NIC2 observations with the
same limitations of resolution and depth.  We find that our simulated
merger remnants occupy a different region in ($n$, $r_e$)-space than
the vD08 quiescent galaxies.  So do simulations of disk galaxies that
are left to evolve in isolation.  The high Sersic indices ($n \gg 4$)
inferred from the mock observations are driven by both the central
cusp (consisting of young stars) and the extended wings (consisting of
old stars).  We explored a number of variations in progenitor
properties that influence the amplitude of the central cusp, but have
little effect on the outskirts.  Subsequent (minor) merging tends to
build up the profile wings even further while leaving the central cusp
in the mass distribution relatively unaffected.

By lack of cosmological context, the binary merger simulations start
with large amounts of gas in the progenitor disks (up to 80\% of the
baryonic mass content) in order to have a sufficiently large gas
fraction by final coalescence to model the dissipational merger event.
Given this large gas reservoir at the start of the simulation, it is
hard to prevent it from forming stars and ending up forming the
extended wings of the remnant profiles (wings that are not seen to
that degree in the observations).  As a consequence, while gas-rich
mergers as simulated in this paper reproduce the basic structural
diagnostic (the size-mass diagram), second order structural properties
such as profile shape may differ because too many stars are formed at
early times.  This could be prevented if the gas was not all present
initially but instead accreted more gradually over time, or if star
formation and/or stellar mass loss and its feedback operated in such a
way that the efficiency of converting gas into stars was lower at low
densities than assumed here (i.e., in the progenitor disks and
particularly their outer parts).  One could speculate that a low
gas-phase metallicity, turbulence from accretion (Bournaud et
al. 2010), or the elevated intensity of the ionizing background
radiation at $z \sim 2$ inhibit the formation of molecular gas and
thus stars during this early phase, especially at low densities where
the gas may still be pristine.

We demonstrate that, in addition to spatial resolution, timestep
resolution is important in modeling the detailed structure of such
extreme systems properly.  Therefore, running spheroid evolution
simulations in a full cosmological context is computationally very
expensive.  Pioneering efforts to simulate spheroid formation in a
cosmological context (without AGN feedback) have recently been carried
out by Naab et al. (2007; 2009).  These authors apply a resimulation
technique on three target halos extracted from a low-resolution dark
matter simulation to follow the formation and evolution of spheroids
within those halos from high redshift to the present day.  In
agreement with our findings, Naab et al. (2007; 2009) argue that at
high redshift dissipational processes on short timescales are a
plausible mechanism to produce the compact galaxies observed at $z
\sim 2$.  At later times, they find that minor and gas-poor merging
forms an efficient means to build up an envelope around this core (see
also Section\ \ref{subsequent.sec} of this paper, and Hopkins et
al. 2009a).\footnote[4]{In detail, as noted already by Naab et
al. (2009), their simulated spheroid with size $r_e = 2.4 \pm 0.4$ kpc
and mass $M = 1.5 \times 10^{11}\ M_{\sun}$ lies in the low tail of
the size distribution of observed nearby galaxies with similar mass.}

In the scenario investigated in this paper, the initial dissipational
event occurs in the form of a major merger.  Such a scenario connects
three types of galaxies observed at high redshift in one evolutionary
sequence: gas-rich disks, dusty starbursts with star formation rates
peaking to $\sim 1000\ M_{\sun}/{\rm yr}$, and quiescent remnants.
The dissipational event modeled by Naab et al. (2007; 2009) is driven
by efficient radiative cooling and collapsing of gas, not necessarily
involving major merging.  It remains to be seen whether such a model
can account for the peak star formation rates observed in, e.g.,
sub-millimeter galaxies, and the post-starburst nature of the stellar
populations of compact galaxies.  Observationally, measurements of
clustering, from large scales to pair counts, based on wide and deep
cosmological surveys could potentially provide support for a
merger-driven evolutionary scenario.

S. Wuyts and P. Jonsson gratefully acknowledge support from the
W. M. Keck Foundation.

\begin{references}
{\small
\reference{} Baker, A. J., Tacconi, L. J., Genzel, R., Lehnert, M. D.,\& Lutz, D. 2004, ApJ, 604, 125
\reference{} Barnes, J. E., Hernquist, L. 1991, ApJ, 370, L65
\reference{} Barnes, J. E. 1992, ApJ, 393, 484
\reference{} Barnes, J. E., Hernquist, L. 1996, ApJ, 471, 115
\reference{} Bezanson, R., van Dokkum, P. G., Tal, T., Marchesini, D., Kriek, M., Franx, M.,\& Coppi, P. 2009, ApJ, 697, 1290
\reference{} Binney, J. 1978, MNRAS, 183, 501
\reference{} Binney, J.,\& Merrifield, M. 1998, Galactic Astronomy, Princeton University Press, Princeton
\reference{} Bournaud, F., et al. 2010, submitted to ApJL (astro-ph\ 1006.4782)
\reference{} Bruzual, G,\& Charlot, S. 2003, MNRAS, 344, 1000
\reference{} Buitrago, F., Trujillo, I., Conselice, C. J., Bouwens, R. J., Dickinson, M.,\& Yan, H. 2008, ApJ, 687, L61
\reference{} Bullock, J. S., Kolatt, T. S., Sigad, Y., Somerville, R. S., Kravtsov, A. V., Klypin, A. A., Primack, J. R.,\& Dekel, A. 2001, MNRAS, 321, 559
\reference{} Cappellari, M., et al. 2006, MNRAS, 366, 1126
\reference{} Cappellari, M., et al. 2009, ApJ, 704, L34
\reference{} Carrasco, E. R., Conselice, C. J.,\& Trujillo, I. 2010, MNRAS, in press (astro-ph\ 1003.1956)
\reference{} Cenarro, A. J.,\& Trujillo, I. 2009, ApJ, 696, L43
\reference{} Chabrier, G. 2003, PASP, 115, 763
\reference{} Cimatti, A., et al. 2008, A\&A, 482, 21
\reference{} Coppin, K. E. K., et al. 2007, ApJ, 665, 936
\reference{} Cox, T. J., Dutta, S. N., Di Matteo, T., Hernquist, L., Hopkins, P. F., Robertson, B.,\& Springel, V. 2006, ApJ, 650, 791
\reference{} Daddi, E., et al. 2005, ApJ, 626, 680
\reference{} Daddi, E., Dannerbauer, H., Elbaz, D., Dickinson, M., Morrison, G., Stern, D.,\& Ravindranath, S. 2008, ApJ, 673, L21
\reference{} Dav\'{e}, R. 2008, MNRAS, 385, 147
\reference{} Dekel, A.,\& Birnboim, Y. 2006, MNRAS, 368, 2
\reference{} Dekel, A., et al. 2009a, Nature, 457, 451
\reference{} Dekel, A., Sari, Re'em,\& Ceverino, D. 2009b, ApJ, 703, 785
\reference{} Di Matteo, T., Springel, V.,\& Hernquist, L. 2005, Nature, 433, 604
\reference{} Di Matteo, P., Pipino, A., Lehnert, M. D., Combes, F.,\& Semelin, B. 2009, A\&A, 499, 427
\reference{} Elmegreen, B. G., Bournaud, F., Elmegreen, D. M. 2008, ApJ, 688, 67
\reference{} Erb, D. K., Steidel, C. C., Shapley, A. E., Pettini, M., Reddy, N. A.,\& Adelberger, K. L. 2006, ApJ, 646, 107
\reference{} Fan, L., Lapi, A., De Zotti, G.,\& Danese, L. 2008, ApJ, 689, L101
\reference{} F\"{o}rster Schreiber, N. M., et al. 2009, ApJ, 706, 1364
\reference{} Franx, M., van Dokkum, P. G., F\"{o}rster Schreiber, N. M., Wuyts, S., Labb\'{e}, I.,\& Toft, S. 2008, ApJ, 688, 770
\reference{} Genzel, R., et al. 2008, ApJ, 687, 59
\reference{} Groves, B., Dopita, M. A., Sutherland, R. S., Kewley, L. J., Fischera, J., Leitherer, C., Brandl, B.,\& van Breugel, W. 2008, ApJS, 176, 438
\reference{} Guo, Q. \& White, S. D. M. 2008, MNRAS, 384, 2
\reference{} Jonsson, P. 2006, MNRAS, 372, 2
\reference{} Jonsson, P., Groves, B.,\& Cox, T. J. 2010, MNRAS, 403, 17
\reference{} Hernquist, L. 1990, ApJ, 356, 359
\reference{} Hopkins, P. F., Hernquist, L., Martini, P., Cox, T. J., Robertson, B., Di Matteo, T.,\& Springel, V. 2005, ApJ, 625, L71
\reference{} Hopkins, P. F., Hernquist, L., Cox, T. J., Di Matteo, T., Robertson, B.,\& Springel, V. 2006, ApJS, 163, 1
\reference{} Hopkins, P. F., Hernquist, L., Cox, T. J., Keres, D. 2008a, ApJS, 175, 356
\reference{} Hopkins, P. F., Hernquist, L., Cox, T. J., Dutta, S. N.,\& Rothberg, B. 2008b, ApJ, 679, 156
\reference{} Hopkins, P. F., Cox, T. J.,\& Hernquist, L. 2008, ApJ, 689, 17 
\reference{} Hopkins, P. F., Hernquist, L., Cox, T. J., Keres, D.,\& Wuyts, S. 2009a, ApJ, 691, 1424
\reference{} Hopkins, P. F., Cox, T. J., Dutta, S. N., Hernquist, L., Kormendy, J.,\& Lauer, T. R. 2009b, ApJS, 181, 135
\reference{} Hopkins, P. F., Lauer, T. R., Cox, T. J., Hernquist, L.,\& Kormendy, J. 2009c, ApJS, 181, 486
\reference{} Hopkins, P. F., Bundy, K., Murray, N., Quataert, E., Lauer, T. R.,\& Ma, C.-P. 2009d, MNRAS, 398, 898
\reference{} Hopkins, P. F., Bundy, K., Hernquist, L., Wuyts, S.,\& Cox, T. J. 2010a, MNRAS, 401, 1099
\reference{} Hopkins, P. F., et al. 2010b, ApJ, 715, 202  
\reference{} Hopkins, P. F.,\& Hernquist, L. 2010, MNRAS, in press (astro-ph\ 1006.0488)
\reference{} Keres, D., Katz, N., Weinberg, D. H.,\& Dav\'{e}, R. 2005, 363, 2
\reference{} Khochfar, S.,\& Silk, J. 2006a, ApJ, 648, L21
\reference{} Khochfar, S.,\& Silk, J. 2006b, MNRAS, 370, 902
\reference{} Kriek, M., et al. 2006, ApJ, 649, 71
\reference{} Kroupa, P. 2001, MNRAS, 322, 231
\reference{} Kuntschner, H., et al. 2010, MNRAS, in press (astro-ph\ 1006.1574)
\reference{} Labb\'{e}, I., et al. 2005, ApJ, 624, L81
\reference{} Lotz, J. M., Jonsson, P., Cox, T. J.,\& Primack, J. R. 2010a, MNRAS, 404, 575
\reference{} Lotz, J. M., Jonsson, P., Cox, T. J.,\& Primack, J. R. 2010b, MNRAS, 404, 590
\reference{} Maraston, C. 2005, MNRAS, 362, 799
\reference{} Maraston, C., Daddi, E., Renzini, A., Cimatti, A., Dickinson, M., Papovich, C., Pasquali, A.,\& Pirzkal, N. 2006, ApJ, 652, 85
\reference{} Mihos, J. C., Hernquist, L. 1994, ApJ, 431, L9
\reference{} Mihos, J. C., Hernquist, L. 1996, ApJ, 464, 641
\reference{} Mo, H. J., Mao, S.,\& White, S. D. M. 1998, MNRAS, 295, 319
\reference{} Muzzin, A., van Dokkum, P. G., Franx, M., Marchesini, D., Kriek, M.,\& Labb\'{e}, I. 2009a, ApJ, 706, 188
\reference{} Muzzin, A., Marchesini, D., van Dokkum, P. G., Labb\'{e}, I., Kriek, M.,\& Franx, M. 2009b, ApJ, 701, 1839
\reference{} Naab, T., Jesseit, R.,\& Burkert, A. 2006, MNRAS, 372, 839
\reference{} Naab, T., Johansson, P. H., Ostriker, J. P.,\& Efstathiou, G. 2007, ApJ, 658, 710
\reference{} Naab, T., Johansson, P. H.,\& Ostriker, J. P. 2009, ApJ, 699, 178
\reference{} Narayanan, D., et al. 2009, MNRAS, in press (astro-ph\ 0910.2234)
\reference{} Narayanan, D., Hayward, C. C., Cox, T. J., Hernquist, L., Jonsson, P., Younger, J. D.,\& Groves, B. 2010, MNRAS, 401, 1613
\reference{} Ocvirk, P., Pichon, C.,\& Teyssier, R. 2008, MNRAS, 390, 13260
\reference{} Paturel, G., et al. 2003, A\&A, 412, 45
\reference{} Peng, C. Y., Ho, L. C., Impey, C. D.,\& Rix, H.-W. 2002, AJ, 124, 266
\reference{} Peirani, S., Crockett, R. M., Geen, S., Khochfar, S., Kaviraj, S.,\& Silk, J. 2010, MNRAS, 405, 2327
\reference{} Rawle, T. D., Smith, R. J.,\& Lucey, J. R. 2010, MNRAS, 401, 852
\reference{} Robertson, B., Cox, T. J., Hernquist, L., Franx, M., Hopkins, P. F., Martini, P.,\& Springel, V. 2006, ApJ, 641, 21
\reference{} Robertson, B., Li, Y., Cox, T. J., Hernquist, L.,\& Hopkins, P. F. 2007, ApJ, 667, 60
\reference{} Ryan, R. E., et al. 2010, submitted to ApJ (astro-ph\ 1007.1460) 
\reference{} Sersic, J. L. 1968, Atlas de galaxias australes (Cordoba, Argentina: Observatorio Astronomico, 1968)
\reference{} Shen, S., Mo, H. J., White, S. D. M., Blanton, M. R., Kauffmann, G., Voges, W., Brinkmann, J.,\& Csabai, I. 2003, MNRAS, 343, 978
\reference{} Springel, V.,\& Hernquist, L. 2002, MNRAS, 333, 649
\reference{} Springel, V.,\& Hernquist, L. 2003, MNRAS, 339, 289
\reference{} Springel, V. 2005, MNRAS, 364, 1105
\reference{} Springel, V., Di Matteo, T.,\& Hernquist, L. 2005, MNRAS, 361, 776
\reference{} Szomoru, D., et al. 2010, ApJ, 714, L244
\reference{} Tacconi, L. J., et al. 2010, Nature, 463, 781
\reference{} Toft, S., et al. 2007, ApJ, 671, 285
\reference{} Tormen, G., Bouchet, F. R.,\& White, S. D. M., 1997, MNRAS, 286, 865
\reference{} Trujillo, I., et al. 2006, ApJ, 650, 18
\reference{} van der Wel, A., Holden, B. P., Zirm, A. W., Franx, M., Rettura, A., Illingworth, G. D.,\& Ford, H. C. 2008, ApJ, 688, 48
\reference{} van der Wel, A.,\& van der Marel, R. P. 2008, ApJ, 684, 260
\reference{} van Dokkum, P. G.,\& Stanford, S. A. 2003, ApJ, 585, 78
\reference{} van Dokkum, P. G., et al. 2006, ApJ, 638, 59
\reference{} van Dokkum, P. G. 2008, ApJ, 674, 29
\reference{} van Dokkum, P. G., et al. 2008, ApJ, 677, 5
\reference{} van Dokkum, P. G., Kriek, M.,\& Franx, M. 2009, Nature, 460, 717
\reference{} van Dokkum, P. G., et al. 2010, ApJ, 709, 1018
\reference{} Vitvitska, M., Klypin, A. A., Kravtsov, A. V., Wechsler, R. H., Primack, J. R.,\& Bullock, J. S. 2002, ApJ, 581, 799
\reference{} Wilkins, S. M., Hopkins, A. M., Trentham, N.,\& Tojeiro, R. 2008, MNRAS, 391, 363
\reference{} Williams, R. J., Quadri, R. F., Franx, M., van Dokkum, P., Toft, S., Kriek, M.,\& Labb\'{e}, I. 2010, ApJ, 713, 738
\reference{} Wuyts, S., et al. 2007, ApJ, 655, 51
\reference{} Wuyts, S., Franx, M., Cox, T. J., Hernquist, L., Hopkins, P. F., Robertson, B. E.,\& van Dokkum, P. G. 2009a, ApJ, 696, 348
\reference{} Wuyts, S., et al. 2009b, ApJ, 700, 799
\reference{} Younger, J. D., Hayward, C. C., Narayanan, D., Cox, T. J., Hernquist, L.,\& Jonsson, P. 2009, MNRAS, 396, 66
\reference{} Zirm, A. W., et al. 2007, ApJ, 656, 66
}
\end {references}


\end {document}